\documentclass{article}
\pdfoutput=1

\PassOptionsToPackage{round}{natbib}

\usepackage[preprint]{neurips_2020}

\usepackage{hyperref}       
\usepackage{color}
\usepackage{url}            
\usepackage{amsfonts, amsmath, amssymb, amsthm}
\usepackage{nicefrac}       
\usepackage{microtype}      
\usepackage{graphicx}				
\usepackage{subcaption}			

\newcommand{\highlightvariables}{0} 

\if1\highlightvariables 
\newcommand{\todo}[1]{\textcolor{red}{TODO: #1}}
\newcommand{\yd}[1]{\textcolor{blue}{YD: #1}}
\newcommand{\lxq}[1]{\textcolor{green}{LXQ: #1}}
\newcommand{\jl}[1]{\textcolor{cyan}{JL: #1}}
\else
\newcommand{\todo}[1]{\ignorespaces}
\newcommand{\yd}[1]{\ignorespaces}
\newcommand{\lxq}[1]{\ignorespaces}
\newcommand{\jl}[1]{\ignorespaces}
\fi

\if1\highlightvariables
\def\draftcolor{magenta}  
\else
\def\draftcolor{black}  
\fi

\newcommand{\githubDANA}{\textcolor{\draftcolor}{\url{https://github.com/LXQin/DANA}}}
\newcommand{\githubSupplements}{\textcolor{\draftcolor}{\url{https://github.com/LXQin/DANA-paper-supplementary-materials}}}

\newcommand{\rversion}{\textcolor{\draftcolor}{4.0.2}}

\newcommand{\inv}[1]{ #1 ^{-1}}  

\newcommand{\concCorr}[2]{\textcolor{\draftcolor}{\ensuremath{\operatorname{CCC}[ #1, #2 ]}}}

\def\R{\mathbb{R}}

\newcommand{\Rpp}{\ensuremath{\R^{p\times p}}}

\newcommand{\method}{\textcolor{\draftcolor}{\texttt{DANA}}}
\newcommand{\mscr}{\textcolor{\draftcolor}{\ensuremath{\texttt{mscr}^{-}}}}
\newcommand{\cc}{\textcolor{\draftcolor}{\ensuremath{\texttt{cc}^{+}}}}

\newcommand{\benchmarkdata}{\textcolor{\draftcolor}{benchmark}}
\newcommand{\testdata}{\textcolor{\draftcolor}{test}}
\newcommand{\singlebatchdata}{\textcolor{\draftcolor}{single-batch}}
\newcommand{\mixedbatchdata}{\textcolor{\draftcolor}{mixed-batch}}

\newcommand{\tzero}{\textcolor{\draftcolor}{\ensuremath{\ell^{-}}}}
\newcommand{\tpoor}{\textcolor{\draftcolor}{\ensuremath{u^{-}}}}
\newcommand{\twell}{\textcolor{\draftcolor}{\ensuremath{\ell^{+}}}}

\newcommand{\poscontrol}{\textcolor{\draftcolor}{\ensuremath{\mathcal{C^{+}}}}}
\newcommand{\negcontrol}{\textcolor{\draftcolor}{\ensuremath{\mathcal{C^{-}}}}}
\newcommand{\numpos}{\textcolor{\draftcolor}{\ensuremath{p^{+}}}}
\newcommand{\numneg}{\textcolor{\draftcolor}{\ensuremath{p^{-}}}}

\newcommand{\poscontrolcluster}{\textcolor{\draftcolor}{\ensuremath{\mathfrak{C}^{+}}}}

\newcommand{\normop}{\textcolor{\draftcolor}{\ensuremath{\operatorname{norm}}}}

\newcommand{\precisionmat}{\textcolor{\draftcolor}{\ensuremath{\Theta}}}
\newcommand{\precision}{\textcolor{\draftcolor}{\ensuremath{\theta}}}
\newcommand{\covariancemat}{\textcolor{\draftcolor}{\ensuremath{\Sigma}}}

\newcommand{\pcor}{\textcolor{\draftcolor}{\ensuremath{\rho}}}
\newcommand{\corrpos}[1]{\textcolor{\draftcolor}{\ensuremath{\pcor^{+}_{ #1 }}}}
\newcommand{\corrposnorm}[1]{\textcolor{\draftcolor}{\ensuremath{\pcor^{\normop,+}_{ #1 }}}}
\newcommand{\corrneg}[1]{\textcolor{\draftcolor}{\ensuremath{\pcor^{-}_{ #1 }}}}
\newcommand{\corrnegnorm}[1]{\textcolor{\draftcolor}{\ensuremath{\pcor^{\normop,-}_{ #1 }}}}

\newcommand{\varZero}{\textcolor{\draftcolor}{\ensuremath{\operatorname{v}_0}}}
\newcommand{\varZeroNorm}{\textcolor{\draftcolor}{\ensuremath{\operatorname{v}_0^{\normop}}}}

\newcommand{\msknumgenes}{\textcolor{\draftcolor}{\ensuremath{1033}}}
\newcommand{\msktesttzero}{\textcolor{\draftcolor}{\ensuremath{2}}}
\newcommand{\msktesttpoor}{\textcolor{\draftcolor}{\ensuremath{10}}}
\newcommand{\msktesttwell}{\textcolor{\draftcolor}{\ensuremath{128}}}
\newcommand{\msktestnumpos}{\textcolor{\draftcolor}{\ensuremath{115}}}
\newcommand{\msktestnumneg}{\textcolor{\draftcolor}{\ensuremath{102}}}

\newcommand{\ucecnumsamples}{\textcolor{\draftcolor}{\ensuremath{48}}}
\newcommand{\ucecnumsamplesEND}{\textcolor{\draftcolor}{\ensuremath{22}}}
\newcommand{\ucecnumsamplesSER}{\textcolor{\draftcolor}{\ensuremath{26}}}
\newcommand{\ucecnumgenes}{\textcolor{\draftcolor}{\ensuremath{1848}}}

\newcommand{\ucecmixedtzero}{\textcolor{\draftcolor}{\ensuremath{2}}}
\newcommand{\ucecmixedtpoor}{\textcolor{\draftcolor}{\ensuremath{5}}}
\newcommand{\ucecmixedtwell}{\textcolor{\draftcolor}{\ensuremath{64}}}
\newcommand{\ucecmixednumpos}{\textcolor{\draftcolor}{\ensuremath{110 }}}
\newcommand{\ucecmixednumneg}{\textcolor{\draftcolor}{\ensuremath{112}}}

\newcommand{\tcganumsamples}{\textcolor{\draftcolor}{\ensuremath{223}}}
\newcommand{\tcganumsamplesBRCA}{\textcolor{\draftcolor}{\ensuremath{166}}}
\newcommand{\tcganumsamplesUCS}{\textcolor{\draftcolor}{\ensuremath{57}}}
\newcommand{\tcganumgenes}{\textcolor{\draftcolor}{\ensuremath{1848}}}
\newcommand{\tcgatzero}{\textcolor{\draftcolor}{\ensuremath{2}}}
\newcommand{\tcgatpoor}{\textcolor{\draftcolor}{\ensuremath{5}}}
\newcommand{\tcgatwell}{\textcolor{\draftcolor}{\ensuremath{100}}}
\newcommand{\tcganumpos}{\textcolor{\draftcolor}{\ensuremath{116}}}
\newcommand{\tcganumneg}{\textcolor{\draftcolor}{\ensuremath{91}}}

\def\figpath{./}	

\begin{document}

\title{Depth Normalization of Small RNA Sequencing:	Using Data and Biology to Select a Suitable Method}

\author{%
	Yannick D\"uren \\
	Mathematical Statistics\\
	Ruhr-University Bochum \\
	44801 Bochum, Germany \\
	\texttt{yannick.dueren@rub.de} \\
	\And
	Johannes Lederer \\
	Mathematical Statistics\\
	Ruhr-University Bochum \\
	44801 Bochum, Germany \\
	\texttt{johannes.lederer@rub.de} \\
	\AND
	Li-Xuan Qin \\
	Memorial Sloan Kettering Cancer Center\\
	485 Lexington Avenue \\
	New York, NY 10017 \\
	\texttt{qinl@mskcc.org} 
}

\maketitle

\begin{abstract}
Deep sequencing has become one of the most popular tools for transcriptome profiling in biomedical studies.
While an abundance of computational methods exists for ``normalizing'' sequencing data to remove unwanted between-sample variations due to experimental handling, there is no consensus on which normalization is the most suitable for a given data set.
To address this problem, we developed ``\method''---an approach for assessing the performance of normalization methods for microRNA sequencing data based on biology-motivated and data-driven metrics.
Our approach takes advantage of well-known biological features of microRNAs for their expression pattern and chromosomal clustering to simultaneously assess (1)~how effectively normalization removes handling artifacts, and (2)~how aptly normalization preserves biological signals.
With \method, we confirm that the performance of eight commonly used normalization methods vary widely across different data sets and provide guidance for selecting a suitable method for the data at hand.
Hence, it should be adopted as a routine preprocessing step (preceding normalization) for microRNA sequencing data analysis.
\method\ is implemented in \texttt{R} and publicly available at~\githubDANA.
\end{abstract}

\section{Introduction}
\label{sec:introduction}

Deep sequencing is prone to systematic non-biological artifacts that arise from variations in experimental handling, similar to other genomics technologies such as microarrays~\citep{Leek2010, Consortium2014}. 
Consequently, a critical first step in the analysis of transcriptome sequencing data is to ``normalize'' the data so that data from different sequencing runs are comparable~\citep{Risso2014, Huang2015, Rahman2015}.
One major source of such handling effects comes from the depth of coverage --- defined as the average number of reads per molecule~\citep{Tarazona2011}.

A plethora of analytic methods for depth normalization have been proposed, including methods based on re-scaling~\citep{Mortazavi2008, Robinson2010a, Anders2010, Dillies2013} and methods based on regression~\citep{Leek2014, Risso2014, Zhang2020}. 
Different normalization methods may lead to different analysis results, and no method has been found to work systematically best in studies comparing the performance of these methods~\citep{Bullard2010, Dillies2013, Soneson2013, Li2020a, Qin2020}.
Rather, their performance strongly depends on the data under study~\citep{Consortium2014, Li2014}.
Currently, a method is often chosen by the analyst based on personal preference and convenience rather than objective criteria governed by the data. 

In this study, we introduce a data-driven and biology-motivated approach to objectively guide the selection of depth normalization methods for the data at hand.
We call our novel approach ``DAta-driven Normalization Assessment'' (\method).
\method's goal is to identify a method that maximally removes depth variations due to disparate experimental handling (``handling effects'') while minimally impacting true biological signals (``biological effects''). 
Its overall three-step strategy is to first define two sets of control markers for capturing each of the two types of variations, then compute statistical measures for quantifying these variations and their change before \textit{versus} after normalization, and, lastly, use numeric metrics and graphical tools for summarizing these measures across each set of control markers.

We apply this concept to the sequencing of microRNAs (miRNAs), a prevalent class of small RNAs that play an important regulatory role of gene expression in the cell ~\citep{Ambros2004, Bartel2004}.
Two key biological features of miRNAs are that (i) they tend to be expressed in an on-off manner where only a subset of miRNAs are expressed in a given sample, and (ii) a subset of them can be organized into polycistronic clusters that tend to be co-regulated and hence co-expressed~\citep{Baskerville2005, Landgraf2007, Griffiths-Jones2008, Chaulk2016}. 
To exploit these known features, we first define a set of negative control markers as those miRNAs that are poorly-expressed (that is, markers with low mean expression) reflecting primarily handling effects. 
When handling effects exist in the data, they manifest as high positive correlations among these negative controls, simply due to the shared handling effects. 
We further define a set of positive control markers as the collection of miRNAs that are well-expressed (that is, markers with high mean expression) and belong to polycistronic clusters. 
The shared biological effects lead to high positive correlations among these positive controls regardless of handling effects.
Our \method\ approach does not require any additional study design (such as balanced sample-to-sequencing-batch assignment) or reference data (such as reliable spike-in markers) that is not readily available.
To the best of our knowledge, \method\ is the first approach to achieve a purely data-driven assessment and selection of depth normalization method for miRNA sequencing data.
While this concept can be generalized and applied to other molecules, such as mRNAs, the definition of control markers may not be as straightforward as for miRNAs and will be the subject of future work.

We benchmark the performance of \method\ using a unique pair of miRNA sequencing data sets for the same set of tumor samples, which were previously collected at Memorial Sloan Kettering Cancer Center~\citep{Qin2020}. 
The first data set was sequenced using uniform handling to minimize handling effects and balanced sample-to-library-assignment to cancel out any residual handling effects in group comparison. 
For the same set of samples, a second data set was collected over the years in the order of tumor sample collection and exhibited excessive depth variations.
We use this pair of data sets to validate our definition of control markers, choice of correlation measures and summary metrics, and the performance of the overall \method\ approach.
We further validate and demonstrate the use of our approach by applying it to two data sets from The Cancer Genome Atlas (TCGA).

\section{Materials and Methods}
\label{sec:materials-and-methods}

We first present our three-step \method\ approach, next introduce the paired data sets from MSK for its benchmarking, and then describe two data sets from TCGA for further validation and demonstration.
The \method\ method is implemented in \texttt{R} and open source~(\githubDANA).

\subsection{\method: A data-driven and biology-motivated normalization selection method} 
\label{subsec:normalization-selection-method}

\subsubsection*{Step I. Definition of negative and positive controls} 

\paragraph{Concept}
Following the ideas in~\citet{Qin2016}, we define poorly-expressed markers as those with mean abundance (that is, read count) in an interval~$[\tzero, \tpoor]$ and well-expressed markers as those with mean abundance over a cutoff~\twell.
It is well known that miRNAs tend to be expressed in an on-off manner where only a subset of miRNAs are expressed in a given sample~\citep{Landgraf2007, Lu2005}.
Hence, poorly-expressed markers mainly reflect those markers that were expressed with read counts driven by handling effects.
These poorly-expressed markers serve as negative controls.
It is further known that miRNAs belonging to a shared polycistronic cluster tend to be co-regulated and, hence, co-expressed~\citep{Baskerville2005, Landgraf2007, Griffiths-Jones2008, Chaulk2016}.
Due to this co-expression, we can consequently expect a high, biology-driven correlation between markers located in a mutual polycistronic cluster, regardless of handling effects.
Therefore, we use well-expressed markers that belong to polycistronic clusters with at least two well-expressed members as positive controls reflecting shared biological effects. 
For simplicity and reproducibility, we define miRNAs to be in a mutual polycistronic cluster if their hairpins are separated by less than 10kb on the chromosome.
This corresponds to the cluster definition on miRBase~\citep{Griffiths-Jones2008}.

\paragraph{Cutoff selection}
The cutoffs $\tzero$, $\tpoor$, and $\twell$ should be chosen based on the empirical distribution of the data.
We set~$\tzero$ such that selected poorly-expressed miRNAs show at least mild expression based on the mean read count histogram.
For all data considered in this paper, we observed that $\tzero = 2$ yielded reasonable results and that different cutoff choices only mildly affect the subsequent analysis.
We choose~$\tpoor$ and~$\twell$ such that each control group consists of a sufficient number of markers, where we recommend $75$ to $150$~markers per group for balancing statistical stability and computational convenience.
Note that if the data contains many subtypes or the number of samples is highly unbalanced across subtypes, it might be necessary to take special care during the selection of negative controls.
For example, if a marker is expressed only in a single subtype but non-expressed in all others, it may be erroneously classified as a negative control marker by simply taking the mean expression.
In such cases, we instead recommend selecting negative control markers based on the maximum across subtype means.

Henceforth, we denote the number of positive and negative controls by~$ \numpos $ and~$ \numneg $, respectively.
For ease of notation, we re-index the positive controls by $ \poscontrol := \{1,\dots,\numpos\} $ and the negative controls by $ \negcontrol := \{1,\dots,\numneg\} $.
Furthermore, we denote the set of all pairs of distinct positive control miRNAs $ (i,j) $ that are in a mutual polycistronic cluster by
\begin{equation*}
	\poscontrolcluster := \bigl\{(i,j) \in (\poscontrol)^{2} \enspace | \enspace (i<j) \enspace \text{and} \; (i \text{ and } j \text{ are in a mutual polycistronic cluster}) \bigr\}.
\end{equation*}

\subsubsection*{Step II. Statistical measures of between-marker correlation}

\paragraph{Concept} 
To measure the level of correlation for a pair of control markers, we estimate either their marginal or partial correlations, depending on the control type.
For negative controls, we quantify the overall strength of inter-marker correlations, which are primarily due to shared handling effects, using Pearson marginal correlations.
For positive controls, on the other hand, we capture only the direct co-expression relation for each pair of markers, upon removing any spurious correlation due to co-expression with other positive controls, using partial correlations.
Note that for the computation of empirical correlations, the sample size $n$ must not be overly small.
In our experiments, we have observed that $n\ge 20$ is sufficient for typical miRNA sequencing data.

\paragraph{Statistical definition of partial correlation}
For estimating partial correlations in positive controls, we assume a parametric distribution for the log2-transformed read counts, where by ``log2-transformation,'' we generally refer to the $\log_2(\cdot + 1)$ function so that zero counts are assigned a value of zero after the transformation.
More specifically, we use a multivariate normal distribution as it is widely used for modeling (log2-transformed) sequencing data~\citep{Law2014}, and as it is a standard distribution in statistics with many tools readily available.
Our pipeline also allows for other distributions, such as Poisson or negative binomial for the count data, but the estimation would be much more challenging~\citep{Zhuang2016}.
Under the assumption of normal-distributed data, the Hammersley-Clifford theorem (see for instance~\citet{Lauritzen1996}) establishes a direct link of partial correlations of $ p $ variables to the entries of their precision matrix~$\precisionmat =\inv{\covariancemat} \in \Rpp$, the inverse of the covariance matrix~$\covariancemat \in \Rpp$ of the data.
For any two variables indexed by~$i$ and $j$, respectively, their partial correlation~$\pcor_{i,j}$ follows the relation
\begin{equation*}
	\pcor_{i,j} = -\dfrac{\precision_{i,j}}{\sqrt{\precision_{i,i}\precision_{j,j}}},
\end{equation*}
where~$ \precision_{i,j} $ is the $ (i,j) $-th entry of the precision matrix~$ \precisionmat=(\precision_{i,j})_{1 \le i,j \le p} $. 
Thus, the partial correlation structure can be estimated using well-established statistical methods for precision matrix estimation such as neighborhood selection~\citep{Meinshausen.2006}, graphical lasso~\citep{Friedman.2008}, or FastGGM~\citep{Ren2015}.
In this study, we use the popular neighborhood selection method proposed by~\citet{Meinshausen.2006} and calibrate its tuning parameter using Bayesian Information Criterion.
We, furthermore, compare the results of the chosen method to all aforementioned methods for precision estimation.

\paragraph{Estimation of correlations before and after normalization}
We estimate partial correlations in positive controls and Pearson correlations in negative controls using un-normalized log-counts as well as normalized log-counts for each normalization method under study.
We denote the estimated partial correlations in positive controls for un-normalized log-counts by~$ \corrpos{} := (\corrpos{i, j})_{1 \le i,j \le \numpos}$, and for an arbitrary normalization method by~$ \corrposnorm{} := (\corrposnorm{i, j})_{1 \le i,j \le \numpos}$.
In parallel, we denote Pearson correlations in negative controls by~$ \corrneg{} := (\corrneg{i, j})_{1 \le i,j \le \numneg},$ and~$ \corrnegnorm{} := (\corrnegnorm{i, j})_{1 \le i,j \le \numneg}$, respectively.

\subsubsection*{Step III. Numerical metrics for summarizing correlation across markers}
The performance of a normalization method is evaluated through a comparison of the estimated correlations before \textit{versus} after normalization in positive controls ($ \corrpos{} $ with~$ \corrposnorm{} $) and in negative controls  ($ \corrneg{} $ with~$ \corrnegnorm{} $). 
We introduce two numeric metrics, one for each control type.

\paragraph{A numeric metric for the reduction of handling effects}
An effective normalization method should maximally remove marginal correlations among negative controls, which are most likely caused by shared handling effects.
Ideally, correlations in the normalized data should be centered around zero with low variance.
Therefore, we propose to compute the mean-squared discrepancy of correlations from zero
\begin{equation*}
	\varZero := \dfrac{1}{\numneg(\numneg-1)} \sum_{i=1}^{\numneg} \sum_{\substack{j=1\\ j\neq i}}^{\numneg}  (\corrneg{i,j}-0)^2,
\end{equation*}
for un-normalized and, analogously, $\varZeroNorm$ for normalized data.
We use the metric
\begin{equation}
	\mscr := \dfrac{\varZero - \varZeroNorm}{\varZero},
\end{equation}
that is, the relative difference of the variance of correlations among negative controls before and after normalization to quantify the reduction of handling effects through normalization.
We call this metric the ``mean-squared correlation reduction in negative controls,'' where a high \mscr\ indicates an effective removal of handling effects in the data.

\paragraph{A numeric metric for the preservation of biological effects}
For each pair of positive control miRNAs that are located in a mutual polycistronic cluster, an ideal normalization method should minimally affect their co-expression.
Mathematically speaking, this means that the relationship between~$\corrpos{i,j}$ and~$\corrposnorm{i,j}$ for all same-cluster positive controls~$(i,j)\in\poscontrolcluster$ should be linear with slope~$1$.
As such, we measure this agreement by the concordance correlation coefficient~\citep{Lin1989}.
We denote the vector of correlations in \poscontrolcluster\ by $ (\corrpos{})_{\poscontrolcluster} $ for un-normalized data, and by $ (\corrposnorm{})_{\poscontrolcluster} $ for normalized data.
We further denote the concordance correlations coefficient of two covariates $ x_1, x_2$ by  $ \concCorr{x_1}{x_2} $.
We use the metric 
\begin{equation}
	\cc\ := \concCorr{(\corrpos{})_{\poscontrolcluster}}{(\corrposnorm{})_{\poscontrolcluster}},
\end{equation}
that is, the ``concordance correlation coefficient of the within-cluster partial correlations among positive controls before and after normalization,'' to quantify the preservation of biological signals.
A close-to-1 \cc\ indicates an apt preservation of biological signals in the data.

\paragraph{Guidelines for method selection}

An optimal normalization maximally removes handling effects (high \mscr) while keeping biological signals intact (\cc\ close to 1).
However, in most cases, there is no clear ``best'' method with maximal~\mscr\ \textit{and} maximal~\cc.
Therefore, one should aim for the best possible trade-off between the proposed statistics for negative and positive controls with an emphasis on keeping biological signals intact (\cc\ close to 1).
The two metrics can be easily assessed by plotting the metrics in a scatter plot for each normalization method under study, where a preferable method should be located towards the top-right quadrant of the plot.

\subsection{Graphical tools}
\label{subsec:graphical-tools}

We introduce several graphical tools for examining the definition of positive and negative controls, providing empirical evidence that supports the choice of \method 's correlation measures and summary metrics, and comparing the two metrics among the normalization methods for their performance assessment.
First, a mean count histogram and mean-standard deviation plot should be used to visually examine the un-normalized data and define positive and negative control markers based on the observed count \textit{versus} variation distribution.
Second, we use marginal correlation histograms, partial correlation heatmaps, and partial correlation scatter plots for (i) visually examining the estimated correlations and (ii) further justifying the choice of the summary metrics~\cc\ and~\mscr. 
Lastly, we use scatter plots of \method 's two metrics for comparing the normalization methods under assessment for guiding the method selection.

\paragraph{Mean-standard deviation plot}
Mean-standard deviation plots are scatter plots of marker-specific standard deviations \textit{versus} marker-specific means after log2-transformation.
For miRNA sequencing data, we typically observe close-to-zero standard deviation for non-expressed markers.
For increasing marker-specific mean, the standard deviation typically increases and ``fans out.''
We recommend that for negative controls, the lower bound \tzero\ is tuned such that selected miRNAs reflect a non-zero and heterogeneous distribution of standard deviations capturing miRNAs that reflect handling effects.

\paragraph{Mean count histogram}
After a log2-transformation of the read count data, we compute the mean read count for each miRNA across all samples.
The distribution and range of reads in the data are then visualized using a histogram plot for the mean read abundance.
This way, miRNAs that are essentially non-expressed can be excluded, and poorly- or well-expressed miRNAs can be easily identified.

\paragraph{Marginal correlation histogram}
We compare the frequency of Pearson correlations among negative controls before \textit{versus} after normalization using a histogram.
The distribution of~\corrneg{}\ and~\corrnegnorm{}\ is shown by binning correlation strengths and counting the number of observations in each bin.
Moreover, we use the correlation histogram for comparing correlation strengths between different data sets, both before and after normalization.
An effective normalization method should center the correlations around zero with small variance.

\paragraph{Partial correlation heatmap}

The partial correlation heatmap shows the matrix of pairwise partial correlations for positive controls~\poscontrol\ alongside the polycistronic clustering.
Its upper triangular part shows the strength of the partial correlation, while the lower part indicates whether the pair of miRNAs is located in a mutual polycistronic cluster.
Using partial correlation heatmaps, we visually identify within-cluster correlations in the data and compare them before and after normalization.
The heatmap also allows the comparison of within-cluster correlations with between-cluster correlations, both before and after normalization.

\paragraph{Partial correlation scatter plot}

The concordance of partial correlations in positive controls before normalization~\corrpos{}\ \textit{versus} after normalization~\corrposnorm{}\ is shown using a scatter plot. 
For each pair of distinct, clustered, positive control miRNAs~$(i,j) \in \poscontrolcluster$, their partial correlation~$(\corrpos{i,j}, \corrposnorm{i,j})$ is plotted in a scatter plot.
An ideal normalization method should minimally affect~$\corrpos{i,j}$, and hence, each $(\corrpos{i,j}, \corrposnorm{i,j})$ should ideally lay on a diagonal with slope~$1$.
Combined for all $(i,j) \in \poscontrolcluster$, this concordance is quantified by our metric~\cc.

\subsection{MSK sarcoma data}

\subsubsection*{A pair of data sets for soft tissue sarcoma}
We have previously collected two data sets for the \textit{same set} of tumor samples at MSK.
The tumor samples were 27 myxofibrosarcoma (MXF) samples and 27 pleomorphic malignant fibrous histiocytoma (PMFH) samples, which were all from newly diagnosed, previously untreated, primary tumors collected at MSK between 2000 and 2012.
The first data set was collected using uniform handling to minimize data artifacts and balanced sample-to-library-assignment (via the use of blocking and randomization) to balance any residual artifacts with the tumor groups under comparison. 
For extra quality assurance, we added two pooled samples shared across all libraries and 10~calibrators spiked-in at fixed concentrations.
For the same set of samples, a second data set was collected without making use of such a careful study design, resulting in unwanted depth variations. 
The number of observed miRNAs is \msknumgenes\ and a detailed description of the data was previously reported in~\citet{Qin2020}.
Throughout this paper, we refer to the uniformly-handled data set as the ``\benchmarkdata'' data set and the non-uniformly-handled data set as the ``\testdata'' data set, following the notation used in~\citet{Qin2020}. 
The collection of both data sets is called the ``MSK data sets.''
The benchmark data set serves a two-fold purpose: (i) it provides a baseline reference for assessing the choice of the control markers and their correlation measures, and (ii) it offers a gold standard of differential expression identification (comparing MXF and PMFH) for checking whether the chosen normalization method, which our approach deems optimal for the \testdata\ data, leads to the closest differential expression assessment.

\subsubsection*{Normalization methods assessed in the \testdata\ data set}
Our data-driven normalization selection method generally works for \textit{any} between-sample normalization method that generates non-negative normalized data. 
As proof of concept, we examined eight commonly used normalization methods that were studied in~\citet{Qin2020}, including six scaling-based methods and two regression-based methods. 
A scaling-based method calculates a scaling factor based on the data for each sample and divides its counts by this factor. 
A regression-based method can be non-parametric or parametric: non-parametric methods are based on, for example, a quantile-quantile plot; parametric methods are based on a linear regression, which typically adopts a covariate representing depth variation (by using a known batch variable or deriving a surrogate batch variable from the data) and including this covariate in a regression framework for the analysis of differential expression. 
In this paper, we examined the following popular methods: Total Count (TC)~\citep{Dillies2013}, Upper Quartile (UQ)~\citep{Bullard2010}, Median (Med)~\citep{Dillies2013}, Trimmed Mean of M-values (TMM)~\citep{Robinson2010a}, DEseq~\citep{Anders2010}, PoissonSeq~\citep{Li2012}, Quantile Normalization (QN)~\citep{Bolstad2003}, and Remove Unwanted Variation (RUV)~\citep{Risso2014} with its three variants RUVg, RUVs, and RUVr.
For a description of these methods, we refer readers to~\cite{Qin2020}.
The \method\ approach is applied to the \testdata\ data set to compare the performance of these normalization methods and guide the choice of a most suitable one.

\subsubsection*{Differential expression assessment in both data sets}
Similar to~\citet{Qin2020}, we first assessed evidence of differential expression in the \benchmarkdata\ data (without normalization) using the voom method~\citep{Robinson2010, Law2014} with a p-value cutoff of~0.01, serving as a ``gold standard.''
We then assessed differential expression in the \testdata\ data before and after normalization and compared it with the gold standard. 
This comparison was summarized numerically using the sensitivity (true positive rate) and the positive predictive value ($1 - \text{false discovery rate}$), at the risk of abusing the terminology.

\subsection{TCGA-UCEC data}
\label{subsubsec:TCGA-data}

We further support and validate our approach with two data sets that are partially paired from the TCGA Uterine Corpus Endometrial Carcinoma (TCGA-UCEC) project~\citep{CancerGenomeAtlasResearchNetwork2013}.
We constructed the two data sets so that they each contain \ucecnumsamples~samples, \ucecnumsamplesEND\ of which were of endometrioid subtype (END) and \ucecnumsamplesSER\ of serous subtype (SER), and so that both data sets have 24 samples in common.
The \ucecnumsamples~samples of the first data set were processed in a single batch (batch number 228.63.0), which we will refer to as the ``\singlebatchdata'' data.
The second data set is composed of 24~samples (11~END and 13~SER) from the first data set and 24~samples (11~END and 13~SER) from other batches, which we will refer to as the ``\mixedbatchdata'' data.
Sequencing counts are available for \ucecnumgenes~miRNAs in each sample.
A detailed description of these data sets is deferred to Section~2.1 of the Supplementary Material.

It is reasonable to assume that the \mixedbatchdata\ data set contains significantly higher variation due to handling effects compared to the \singlebatchdata\ data set.
At the same time, due to the sample overlap and identical sample size, biological differences between the two subtypes should be similar between the two data sets.
In other words, the \singlebatchdata\ data can provide a ``silver standard'' for the differential expression status, and the \mixedbatchdata\ data can serve as a test data set for normalization methods and their selection.
We follow the same steps for normalizing the \mixedbatchdata\ data and performing differential expression analysis in each data set as for the MSK data.

\subsection{Combined TCGA-BRCA and TCGA-UCS data}
\label{subsec:BRCA-UCS-data}

We lastly demonstrate the necessity and effectiveness of our \method\ approach in a TCGA data set \textit{without an available benchmark}.
We combined all \tcganumsamplesBRCA~stage III/IV samples from the TCGA Breast Invasive Carcinoma (TCGA-BRCA) project~\citep{CancerGenomeAtlasNetwork2012} with all \tcganumsamplesUCS~samples from the Uterine Carcinosarcoma (TCGA-UCS) project~\citep{Cherniack2017}. 
The total sample size ($n=\tcganumsamples$) is an order of magnitude greater than that for the MSK and TCGA-UCEC data, and \tcganumgenes~miRNAs were observed.
Normalization assessment was applied analogously to the MSK test data and the TCGA-UCEC \mixedbatchdata\ data using the \method\ approach.

\section{Results}
\label{sec:results}

In this section, we present the results of our proposed \method\ approach for the MSK data and TCGA data.
We show that (i) using the suggested statistical tools handling effects can be identified in the data, (ii) normalization is needed, (iii) the relative performance of normalization methods is data-dependent, and (iv) our approach guides the choice for a most suitable method.
All results were generated in \texttt{R} version~\rversion, and the code and data are available on GitHub (\githubSupplements).

\subsection{Normalization assessment for the MSK data sets}
\label{subsec:results-MSK}

\begin{figure}
	\centering
	\includegraphics[width=0.9\textwidth]{\figpath/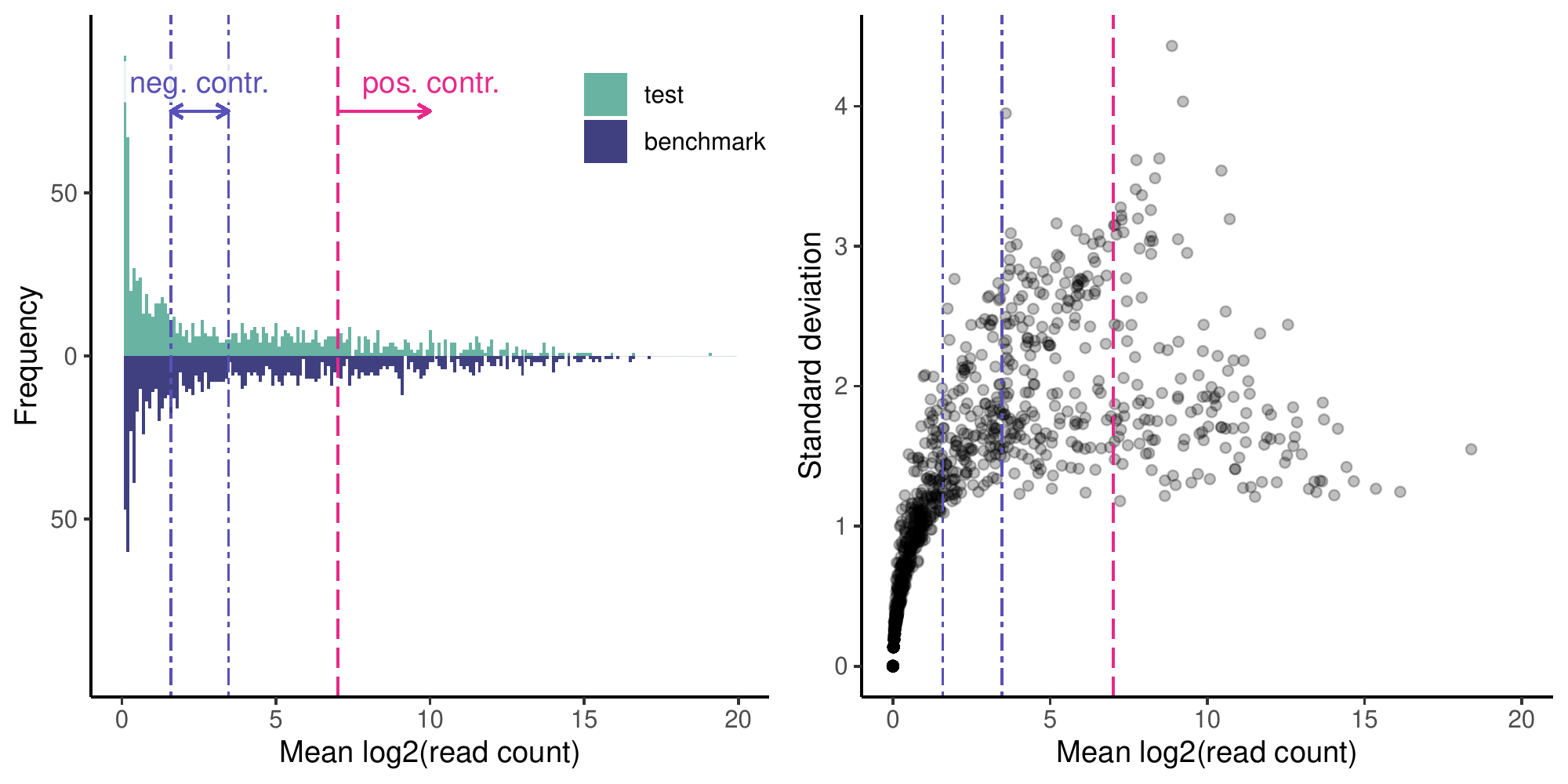} 
	\caption{
		Mean count histogram (left) and mean-standard deviation plot (right) for the MSK \testdata\ data.
		The ranges~$[\tzero,\tpoor]=[\msktesttzero,\msktesttpoor]$ for negative controls and~$[\twell,\infty)=[\msktesttwell,\infty)$ for positive controls are indicated by blue and red vertical lines, respectively.
	}
	\label{fig:MSK-control-choice}
\end{figure}

\begin{figure}
	\centering
	\includegraphics[width=0.6\textwidth]{\figpath/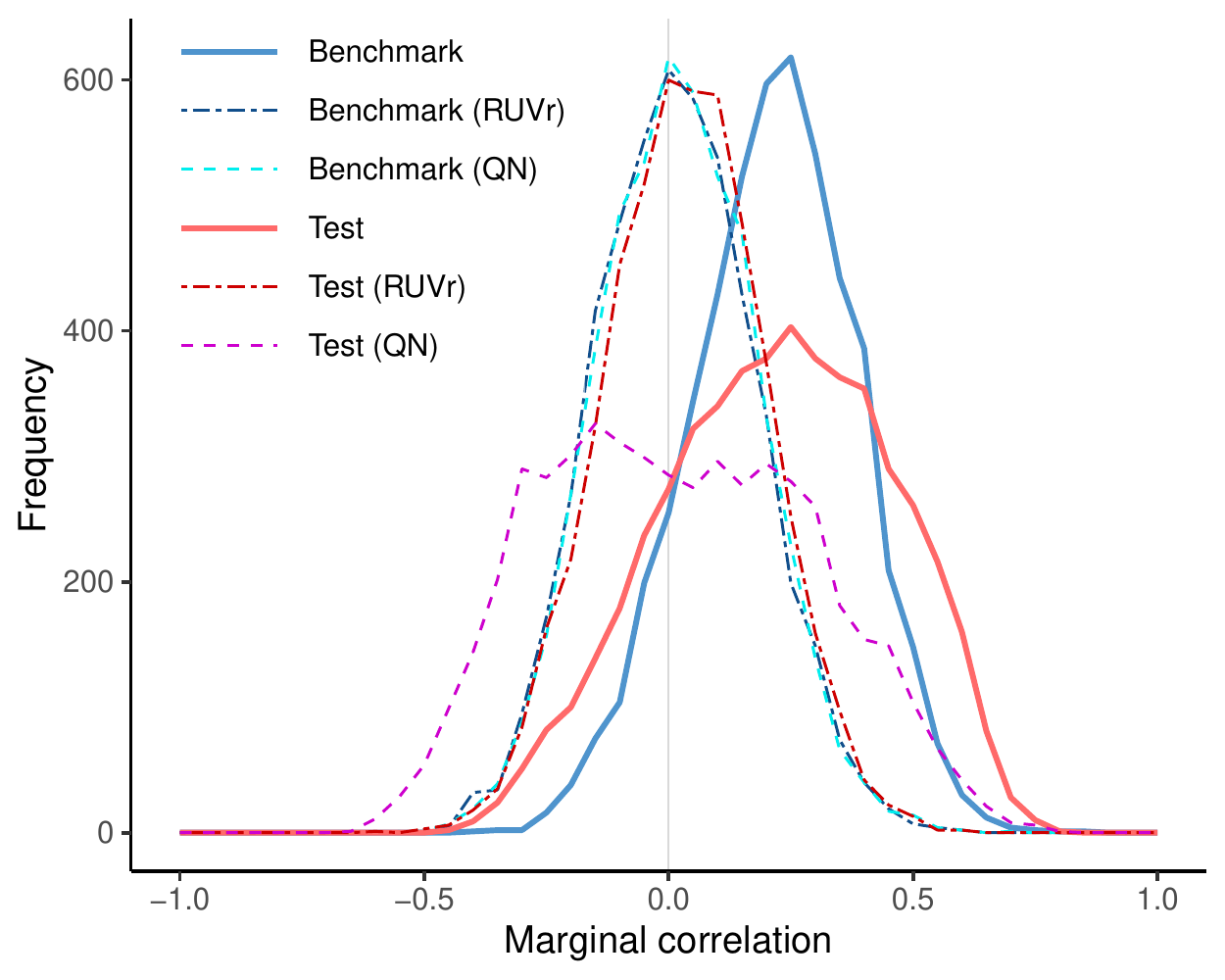} 
	\caption{
		Density curves for Pearson correlations in negative controls. 
		Comparison between the un-normalized, RUVr-normalized, and QN-normalized MSK \benchmarkdata\ data, and the un-normalized, RUVr-normalized, and QN-normalized MSK \testdata\ data.
		The density curves for all normalization methods are shown in Supplementary Figs.~7 and~8.
	}
	\label{fig:MSK-corr-density-neg}
\end{figure}

\subsubsection*{Definition of positive and negative controls and their empirical validation}

We selected the ranges~$[\twell,\infty)$ and~$[\tzero,\tpoor]$ for positive and negative controls, respectively, based on the observed mean count distribution and mean-standard deviation plots for the \testdata\ data set (Fig.~\ref{fig:MSK-control-choice}). 
We set the threshold $ \tzero $ to $ \msktesttzero $ such that (i) non-expressed miRNAs are excluded, (ii) selected miRNAs show at least mild expression, and (iii) selected miRNAs have a heterogeneous distribution of their standard deviation.
We selected the thresholds $\tpoor=\msktesttpoor$ leading to $\numneg=\msktestnumneg$ negative controls, and $\twell=\msktesttwell$ resulting in $\numpos=\msktestnumpos$ positive controls.
The results of the subsequent analyses are robust with respect to the selected ranges, which we show in Section~1.1 of the Supplementary Material, including Supplementary Fig.~1.

\subsubsection*{Empirical justification of the correlation measures and summary metrics}

We provide empirical evidence that supports the choice of correlation measures and the use of the numeric metrics \mscr\ and \cc\ for quantifying the reduction of handling effects and preservation of biological signals, respectively.

For negative controls, we assess the level of their inter-marker correlation in the \testdata\ data before and after normalization, and compare these with that in the \benchmarkdata\ data (Fig.~\ref{fig:MSK-corr-density-neg}). 
First, we observe that the correlations of the un-normalized data sets are not centered around zero, and the mean correlation strength is about 0.21.
This was expected in both data sets due to the nature of high-throughput data, where markers in the same sample are profiled in the same experimental unit.
On the other hand, the variation of correlation strengths in the \testdata\ data is much higher, and, in particular, stronger positive correlations are much more abundant compared to the \benchmarkdata\ data.
This finding accords with our assumption that handling effects manifest themselves as excessive positive correlations and reflects the higher quality of the \benchmarkdata\ data.

Introducing normalization to the data sets, the correlation means are shifted towards zero. 
While the correlation variance in the \benchmarkdata\ data remains approximately equal, it strongly varied for the \testdata\ data depending on the applied normalization.
For example, RUVr normalization leads to a strong variance reduction of inter-marker correlations in the \testdata\ data and a correlation distribution similar to that of the \benchmarkdata\ data.
On the other hand, QN, for example, centers the distribution but considerably increases variance. 
The \mscr\ metric jointly captures both desired normalization effects --- centering around zero and variance reduction --- and, hence, offers a simple and easy-to-interpret metric to assess the reduction of handling effects in the data.
For the sake of clarity, only two exemplary methods are shown in Fig.~\ref{fig:MSK-corr-density-neg}; for all methods, see Supplementary Figs.~7 and~8.

\begin{figure}
	\centering
	\includegraphics[width=0.95\textwidth]{\figpath/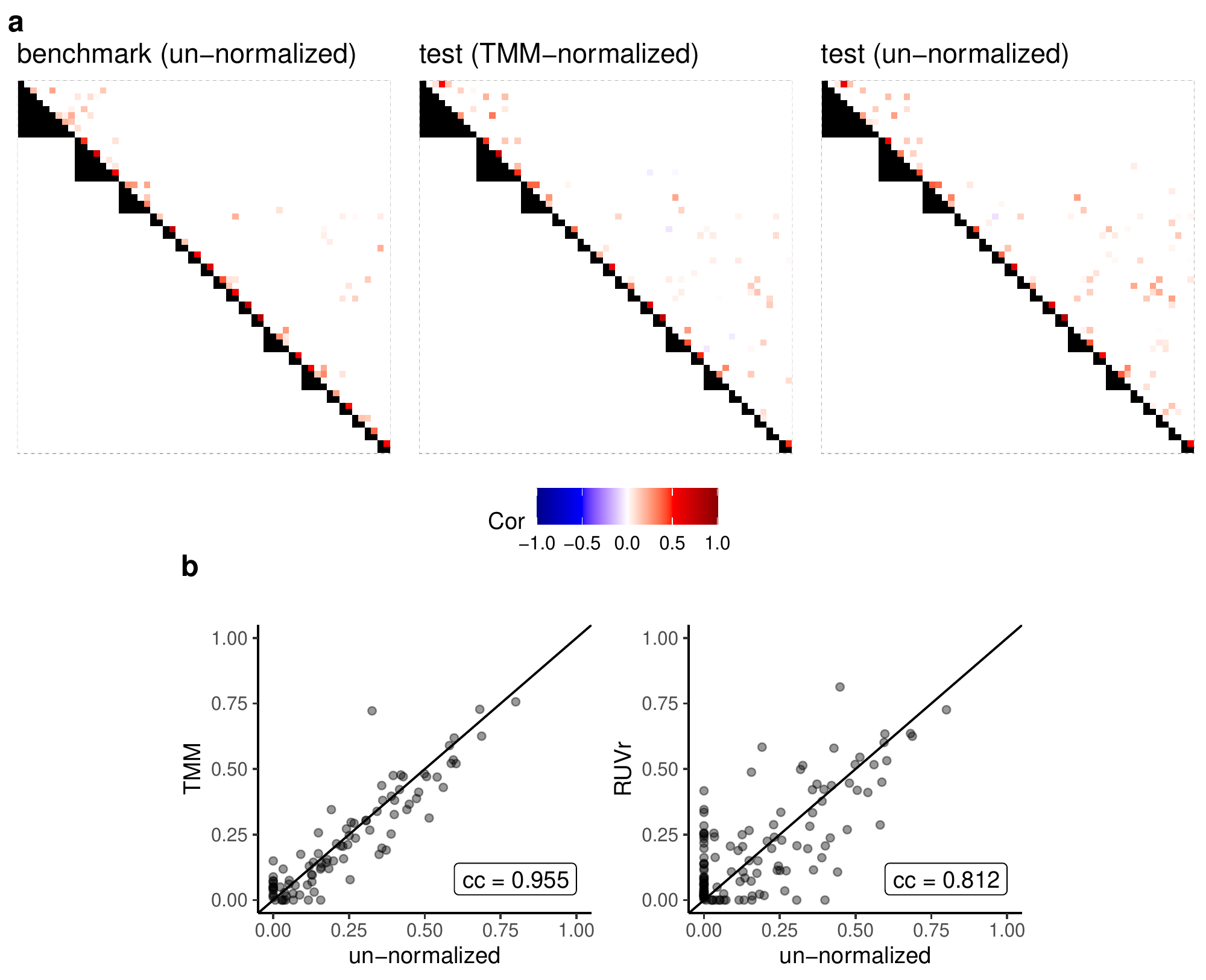} 
	\caption{
		\textbf{a}:~Partial correlation heatmaps for a subset of positive controls in the un-normalized MSK \benchmarkdata\ (left),  TMM-normalized \testdata\ (center), and un-normalized \testdata\ (right) MSK data. 
		Polycistronic clusters are indicated on the lower triangular part.
		\textbf{b}:~Partial correlation scatter plots for within-cluster correlations using TMM normalization (left) and RUVr normalization (right), each in comparison with the un-normalized \testdata\ data.
		The concordance correlation ($\texttt{cc}$) of within-cluster correlations before \textit{versus} after normalization is shown in the bottom right.
		See Supplementary Figs.~5 and~6 for the partial correlation heatmaps and scatter plots, respectively, for all methods.
	}
	\label{fig:MSK-Results-PosControl}
\end{figure}

For positive controls, there is a high abundance of within-cluster correlations in the \benchmarkdata, and all of these correlations are strictly positive (Fig.~\ref{fig:MSK-Results-PosControl}a). 
This aligns well with the biological evidence of co-expression of miRNAs in polycistronic clusters~\citep{Baskerville2005, Landgraf2007, Griffiths-Jones2008, Chaulk2016}.
Compared to the \benchmarkdata, the \testdata\ data has fewer positive within-cluster correlations and more positive off-cluster correlations, both of which likely resulted from excessive handling effects.
Normalization, regardless of the method, alleviated the off-cluster correlations in terms of both the number and strength. 
Depending on the method, normalization can either retain or reduce within-cluster correlations, which is signified by their before-versus-after-normalization concordance. 
The concordance is high for TMM (\cc: 0.955) and low for RUVr (\cc: 0.812) (Fig.~\ref{fig:MSK-Results-PosControl}b). 
Hence, for the MSK \testdata\ data set, TMM better preserves biological effects in positive controls than RUVr.
Note that while RUVr offers the highest reduction of handling effects in negative controls, it over-corrects and fails to preserve biological signals in positive controls.
Hence, it is crucial to assess normalization based on both metrics in tandem.

\subsubsection*{\method\ selection of a suitable normalization method for the MSK \testdata\ data}

\begin{figure}
	\centering
	\includegraphics[width=0.45\textwidth]{\figpath/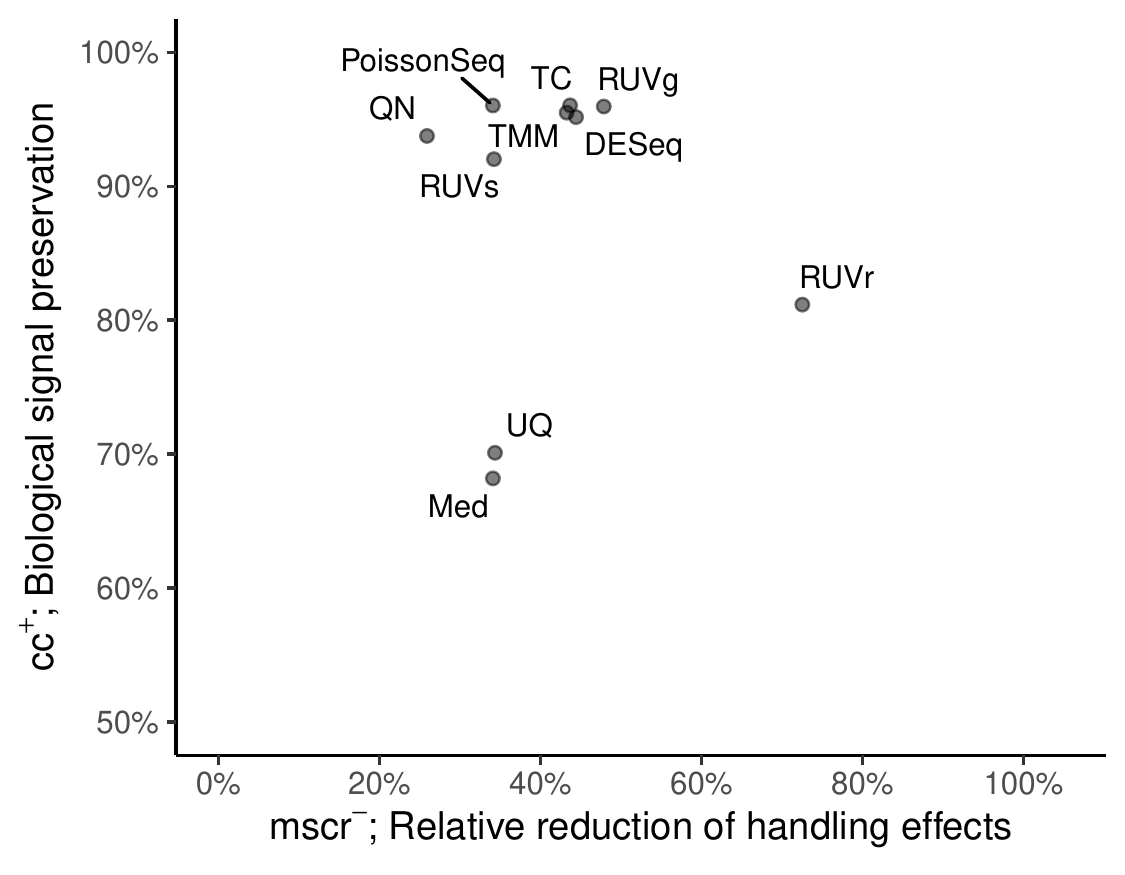}\\
	\caption{
		\method\ summary metrics \mscr\ and \cc\ for the MSK \testdata\ data.
	}
	\label{fig:MSK-result-statistics}
\end{figure}

We computed the \method\ correlation measures for all eight normalization methods applied to the \testdata\ data (Fig.~\ref{fig:MSK-result-statistics}).
RUVr (\cc: 81.2\%; \mscr: 72.53\%) achieved the highest reduction of handling effects, as measured by \mscr, but only moderately preserved biological signals, as measured by \cc. 
RUVg (96.0\%; 47.9\%) offers the best compromise between a very high \cc\ and a high \mscr.
DESeq (95.2\%; 44.4\%), TMM (95.5\%; 43.3\%), and TC (96.1\%; 43.1\%) all performed well with high \cc\ and relatively high \mscr, indicating a good reduction of handling effects and good preservation of biological signals. 
They were followed closely by PoissonSeq (96.0\%; 34.1\%) and RUVs (92.0\%; 34.2\%).
While QN (93.8\%; 25.9\%) kept biological signals mostly intact, it reduced handling effects much less than the aforementioned methods.
The method with the worst \mscr\ was QN (93.8\%; 25.9\%).
Finally, UQ (70.1\%; 34.4\%) and Med (68.2\%; 34.1\%) performed poorly with a mild reduction of handling effects and the overall worst preservation of biological signals.
Hence, for the MSK \testdata\ data, our findings suggest choosing RUVg, TMM, TC, or DESeq for depth normalization and strongly discourage the use of Med, UQ, QN, or RUVr normalization.


The results of our normalization assessment for this data set are in accordance with our prior findings as published in~\citet{Qin2020}; across both studies, the same methods were selected among the best and worst performers, respectively.

\subsubsection*{Comparison of partial correlation estimation methods}

Computation of the \cc~metric requires the estimation of partial correlations among positive controls. 
While we used the popular neighborhood selection method~\citep{Meinshausen.2006} and calibrated its tuning parameter using Bayesian information criterion in this study, we further examined our findings for other partial correlation estimation methods.
We compared the \method\ results for the \testdata\ data using (i) neighborhood selection with different tuning parameter calibration methods, (ii) the glasso method~\citep{Friedman.2008} using various tuning parameter selection methods, and (iii) the FastGGM method~\citep{Ren2015}.
In summary, we found that the \method\ assessment is robust with respect to the chosen precision estimation method and observed a high correlation of the metric~\cc\ across most methods (Supplementary Figs.~3 and~4).
We defer a detailed comparison to Section~1.2 of the Supplementary Material.

\subsection{Normalization assessment for the TCGA-UCEC data}
\label{subsec:results-TCGA-UCEC}

\begin{figure}
	\centering
	\includegraphics[width=0.95\textwidth]{\figpath/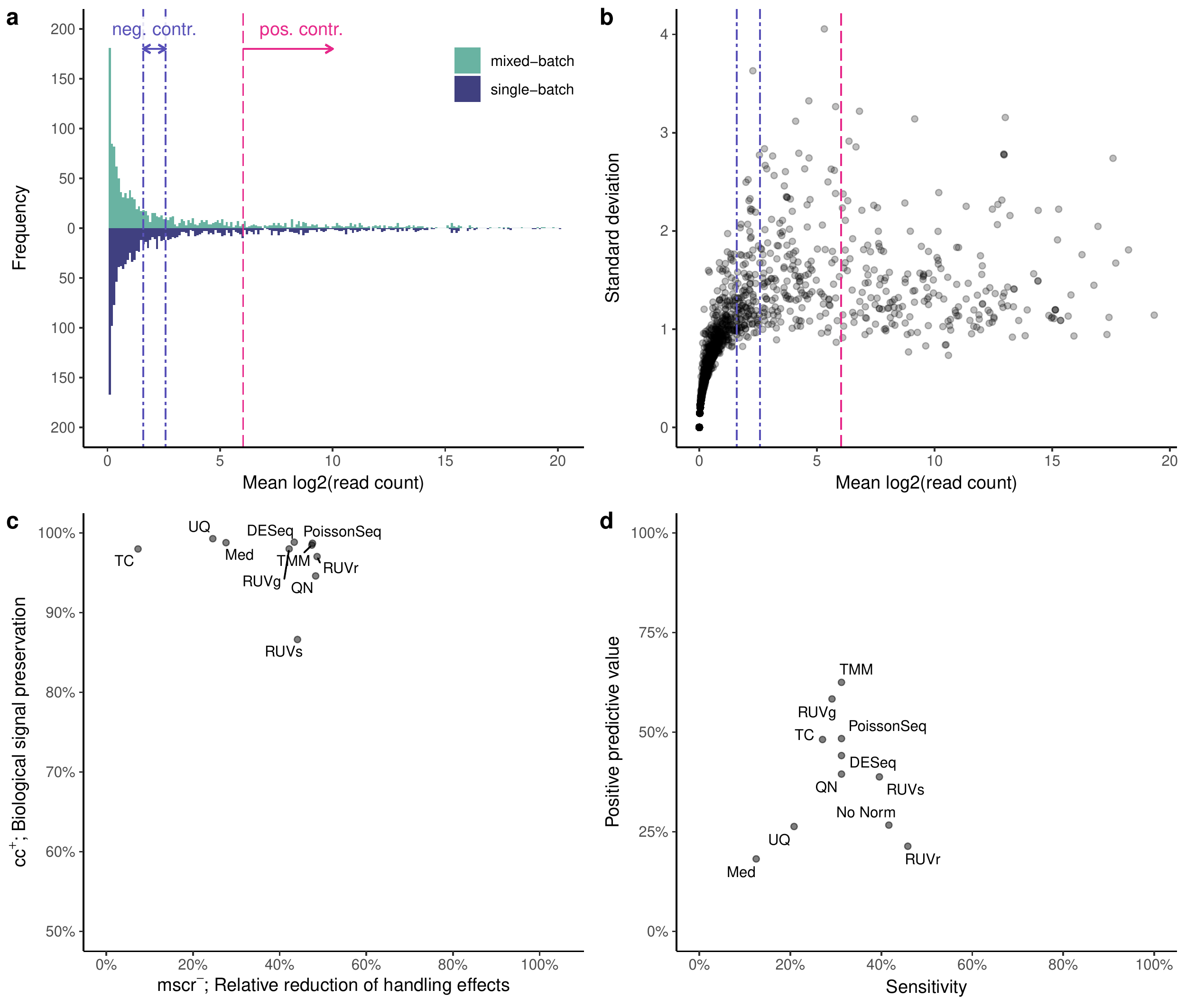} 
	\caption{
		Cutoff selection and normalization assessment for the  TCGA-UCEC endometrial cancer data.
		\textbf{a}:~Mean count histogram plot, and \textbf{b}:~mean-standard deviation plot for log2-transformed TCGA-UCEC data.
		The selected ranges~$[\tzero,\tpoor]=[\ucecmixedtzero,\ucecmixedtpoor]$ for negative controls and~$[\twell,\infty)=[\ucecmixedtwell,\infty)$ for positive controls are indicated by blue and red vertical lines, respectively.
		\textbf{c}:~Normalization assessment using \method\ in the TGCA-UCEC \mixedbatchdata\ data for all normalization methods considered in this study. 
		\textbf{d}:~Normalization assessment based on differential expression analysis between the subtypes END and SER. The differential expression statuses in the \mixedbatchdata\ data (before and after normalization) are compared to those in the \singlebatchdata\ data as an assumed truth and their agreement is summarized using the positive predictive value and the sensitivity.
	}
	\label{fig:UCEC-results}
\end{figure}

For the TCGA-UCEC \mixedbatchdata\ data set, we selected the ranges~$[\tzero,\tpoor]=[\ucecmixedtzero,\ucecmixedtpoor]$ for the definition of negative controls and~$[\twell,\infty)=[\ucecmixedtwell,\infty)$ for positive controls.
This selection was based on the same reasoning we used for the MSK data and facilitated by the same graphical tools (Figs.~\ref{fig:UCEC-results}a,b). 
The number of positive and negative controls for the \mixedbatchdata\ data is $\numpos=\ucecmixednumpos$ and $\numneg=\ucecmixednumneg$, respectively.

All considered normalization methods achieved a very high preservation of biological signals ($ \cc > 94\% $) except RUVs (\cc: 86.6\%; \mscr: 44.1\%) (Fig.~\ref{fig:UCEC-results}c).
However, different methods led to different reduction of handling effects. 
TMM (98.5\%; 47.4\%) and PoissonSeq (98.7\%; 47.6\%) offer a very high \cc combined with a high \mscr, whereas TC (98.0\%; 7.3\%), UQ (99.3\%; 24.6\%), and Med (98.8\%; 27.6\%) performed very poorly in terms of handling effects reduction.
QN (94.6\%; 48.3\%) and RUVr (97.0\%; 48.7\%) have the highest  \mscr\ but sub-optimal \cc.
Hence, TMM and PoissonSeq are most suitable for normalizing the TCGA-UCEC \mixedbatchdata\ data with DESeq (98.8\%; 43.4\%), RUVg (98.0\%; 42.2\%), and RUVr as runners-up.
On the other hand, we discourage using TC, UQ, Med, or RUVs normalization for this data set.
The corresponding marginal correlation histograms, partial correlation heatmaps, and partial correlation scatter plots (Supplementary Figs.~9--11) further confirmed these results, see Section 2.2 of the Supplementary Material.


Among the \ucecnumgenes~miRNAs in the data, differential expression analysis identified 75 to be differentially expressed (DE) in the \singlebatchdata\ data and 48 in the un-normalized mixed-batch data, with a significance cutoff of p-value less than~0.01.
We summarized the agreement of the DE statuses in the \singlebatchdata\ data as an assumed truth to the DE statuses in the normalized and un-normalized \mixedbatchdata\ data using the positive predictive value (PPV, $1 - \text{false discovery rate}$) and the sensitivity (true positive rate) (Fig.~\ref{fig:UCEC-results}d).
Compared to the un-normalized \mixedbatchdata\ (number of DE genes: 48; PPV: 27\%; sensitivity: 42\%), three methods decreased the PPV: Med (33; 18\%; 12\%), UQ (38; 26\%; 21\%), and RUVr (103; 21\%; 46\%); in addition, Med and UQ also decreased sensitivity.
RUVr slightly increased the sensitivity and is the only method that increased the number of DE genes.
The methods TMM (24; 62\%; 31\%), RUVg (24; 58\%; 29\%), PoissonSeq (31; 48\%; 31\%), DESeq (34; 44\% 31\%), TC (27; 48\%; 27\%), RUVs (49; 39\%; 40\%), and QN (38; 39\%; 31\%) all increased PPV but decreased sensitivity.
TMM had the overall highest PPV out of these methods and moderate sensitivity.

Taken together, in our differential expression analysis, TMM tended to outperform the other methods, while Med and UQ were the worst performers. 
Hence, for the TCGA-UCEC data sets, the results of our differential expression study, again, align well with the results of our \method\ approach.


\subsection{Normalization assessment for the combined TCGA-BRCA and TCGA-UCS data}
\label{subsec:results-TCGA}

\begin{figure}
	\centering
	\includegraphics[width=0.45\textwidth]{\figpath/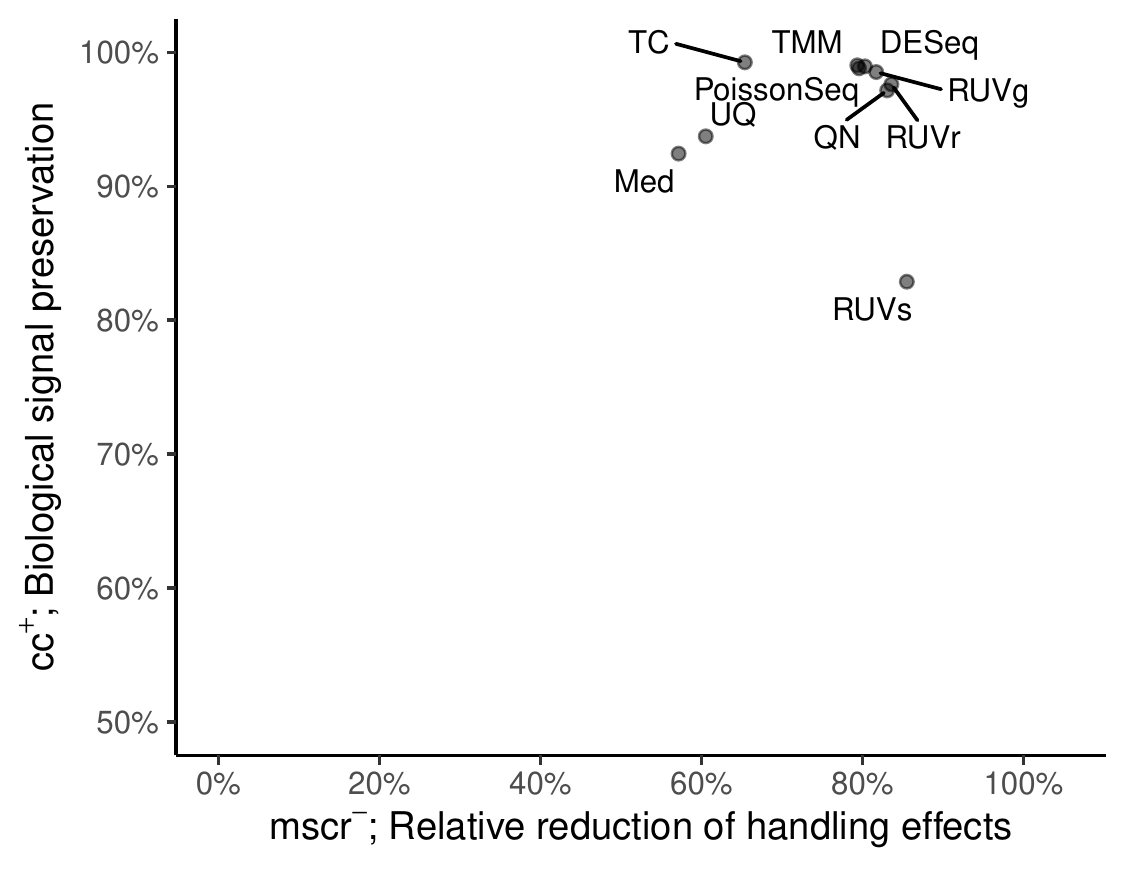}\\
	\caption{
		\method\ summary metrics \mscr\ and \cc\ for the combined TCGA-BRCA and TCGA-UCS data set.
	}
	\label{fig:TCGA-BRCA-UCS-result-metrics}
\end{figure}

Lastly, we demonstrate the effectiveness of the \method\ approach for the combined breast and uterine cancer data set (TCGA-BRCA and TCGA-UCS), for which no gold or silver standard is available.
We selected the ranges~$[\tzero,\tpoor]=[\tcgatzero,\tcgatpoor]$ and~$[\twell,\infty)=[\tcgatwell,\infty)$ based on our recommended data-driven, graphical criteria (Supplementary Fig.~12). 
The number of negative controls is~$\numneg=\tcganumneg$ and the number of positive controls is~$\numpos=\tcganumpos$.

Normalization assessment through \method\ (Fig.~\ref{fig:TCGA-BRCA-UCS-result-metrics} and Supplementary Figs.~13--15) reveals that RUVr and QN offer the best performance for the combined BRCA and UCS data set. 
Both methods have shown high preservation of biological signals as well as a high reduction of handling effects.
As for the TCGA-UCEC data, Med, UQ, and TC again performed worse compared to all other tested normalization methods.
TMM, PoissonSeq, RUVg, and DESeq showed high \cc\ with intermediate \mscr.
RUVs offered the highest reduction of handling effects but showed the worst biological signal preservation, highlighting the necessity of both metrics in our assessment.
In summary, even though no benchmark or silver/gold standard is available, \method\ could effectively assess the performance of normalization methods, and we recommend using RUVr or QN for this particular data set.


\section{Discussion}
\label{sec:discussion}

When done appropriately, normalization can improve the accuracy and reproducibility of subsequent statistical analyses, such as the identification of disease susceptibility genes~\citep{Leek2010, Jaffe2015}. 
However, it has been shown that improper normalization can lead to prominent findings that cannot be reproduced~\citep{Ransohoff2005a, Akey2007, Rahman2015}.
Hence, the selection of a suitable normalization method is a crucial step in transcriptome sequencing data analysis.
Despite numerous publications on the comparison and assessment of normalization methods, there has been no consensus on which normalization method works systematically best for which type of data.
For example, the comparison of 14~studies using normalization methods for downstream differential expression analysis in \citet[Table~1]{Li2020a} and the summary of the literature on normalization comparison in \citet[Table~2]{Evans2018} demonstrate that no method yielded consistently high performance for different data sets.
In this study, we addressed this pressing and under-studied problem by developing a data-driven approach for normalization assessment and selection in miRNA sequencing, where we focused on keeping biological signals in the data intact while effectively removing unwanted variations due to heterogeneous experimental handling.

We based our approach on control markers that are defined by the data and biology, and we catered the choice of numerical measures and summary metrics to each type of control markers. 
Using the carefully benchmarked MSK data, we were able to empirically justify the biology-driven definition of the control markers and validate the choice of numerical measures and summary metrics. 
We have previously successfully used similar definitions of negative controls and positive controls for assessing the impact of normalization in the context of microarray data~\citep{Qin2016}.
However, we acknowledge that DANA would not be suitable for data sets in which the biological variation is expected to be zero across the samples under study for a sizable proportion of the positive control markers.
Finally, we demonstrated that our approach is also robust with respect to the cutoff choices for positive and negative controls.

Across the three empirical studies we conducted in this paper, the performance of normalization methods varied greatly.
A method may perform well for one data set but poorly for others.
For example, based on the \method\ assessment, we recommend using RUVr for the combined TCGA breast and uterine cancer data set; in stark contrast, RUVr performed poorly for the MSK sarcoma cancer data set.
Hence, we confirm the necessity of a \textit{data-driven} selection of a suitable normalization method, which previous studies are lacking.
In this study, we implemented \method---a tool that provides such data-driven guidance for miRNA data.
In our three empirical studies, two methods, upper quartile and median normalization, are consistently among the worst performers based on the two \method\ metrics.
Furthermore, total count normalization, which is nevertheless widely used due to its simplicity, performed poorly to mediocre at best and needs to be reconsidered given our findings.
Note that we do not differentiate between different types of unwanted variation in the data and, consequently, different types of normalization, as they tend not to be clearly distinguished in practice.
For example, TCGA uses only total count normalization and not batch effect correction for pre-processing its miRNA sequencing data, even though their data were collected in multiple batches for each cancer type. 
However, normalization may perform poorly in presence of strong batch effects.
Finally, while it is beyond the scope of this paper, the choice of sequence alignment and feature counting methods likely also affects the performance of normalization.

The application of our \method\ approach does not depend on the availability of benchmark data. 
It applies to any set of miRNA sequencing data, such as those from TCGA. 
For studies with very small sample sizes, however, efforts should be made to profile the data in a single experiment with uniform handling, as \method 's metrics may not be statistically meaningful.
In general, \method\ can assess any normalization method that generates non-negative counts.
While the assumptions of each normalization method should be checked and satisfied by the data, there are many situations in which the validity of any assumption is unknown for the given experiment, or normalization methods can perform well even if their assumptions are (partially) violated~\citep{Evans2018}.
To the best of our knowledge, \method\ is the first approach that provides a purely data-driven assessment of normalization for miRNA sequencing data and does not depend on the availability of reliable spike-ins, housekeeping genes, or a gold standard.

\section{Conclusion}
\label{sec:conclusion}

In this study, we confirmed that normalization assessment is in urgent need to objectively guide the selection of depth normalization methods for the data at hand.
We developed \method---a tool that is data-driven and biology-motivated---for guiding such selection.
Our results also show that there is still a need for more effective normalization methods and support the practice of careful study design, such as uniform handling and a balanced sample-to-library-assignment, for circumventing the need for normalization. %
While we validated our approach using the most common downstream analysis (differential expression analysis) in this paper, we plan to test it using other analysis methods in future work.
Moreover, we will extend the \method\ approach towards application to data generated using other high-throughput profiling platforms.
Finally, we plan to further develop our general approach for other molecules, such as mRNAs, by defining suitable sets of control markers and metrics measuring the degree of preservation of biological effects and removal of handling effects for these molecules. %

\section*{Acknowledgements}
This work was partially supported by the National Institutes of Health [CA214845 to YD and LXQ, CA008748 to LXQ]; and the Deutsche Forschungsgemeinschaft [451920280 to YD and JL].
We thank Joseph Christoff, Shih-Ting Huang, Mike Laszkiewicz, Nils M\"uller, Andy Ni, Nicole Rusk, Mahsa Taheri, Miao Wang, Yilin Wu, Qihang Yang, and Jian Zou for many helpful discussions.


\bibliographystyle{apalike}
\bibliography{library}

\end{document}


\maketitle

This is the supporting material to our paper on \textit{``Depth Normalization of Small RNA Sequencing: Using Data and Biology to Select a Suitable Method''}. 
We show additional results of the \method\ approach for assessing the performance of normalization methods in microRNA sequencing data using (1) the paired MSK data sets, (2) the partially-paired TCGA-UCEC data sets and (3) the combined TCGA-BRCA and TCGA-UCS data set.
All results were generated in \texttt{R} version~\rversion\ and the code, data, and results are available on GitHub at \githubSupplements.



\section{MSK Data}
\label{sec:msk-data}

We further study the paired MSK data sets for the same set of tumor samples for which a detailed description of the data was previously reported in~\citet{Qin2020}.
Recall that one data set (that is, the \benchmarkdata\ data set) was collected using (1) uniform handling to minimize data artifacts and (2) balanced sample-to-library-assignment (via the use of blocking and randomization) to balance any residual artifacts with the tumor groups under comparison.
For the same set of samples, a second data set (that is, the \testdata\ data set) was collected without making use of such a careful study design, resulting in unwanted depth variations. 
Using this pair of data sets, we empirically examine the \method\ approach by evaluating its sensitivity for different control marker choices, assessing different method choices for partial correlation estimation, and comparing control-marker correlations in negative and positive controls between the two data sets.

\subsection{Cutoff choices for defining control makers}
\label{subsec:MSK-cutoff-study}

\newcommand{\configa}{\textcolor{\draftcolor}{(\textbf{a})}}
\newcommand{\configb}{\textcolor{\draftcolor}{(\textbf{b})}}
\newcommand{\configc}{\textcolor{\draftcolor}{(\textbf{c})}}

\begin{figure}
	\centering
	\includegraphics[width=0.99\textwidth]{\figpath/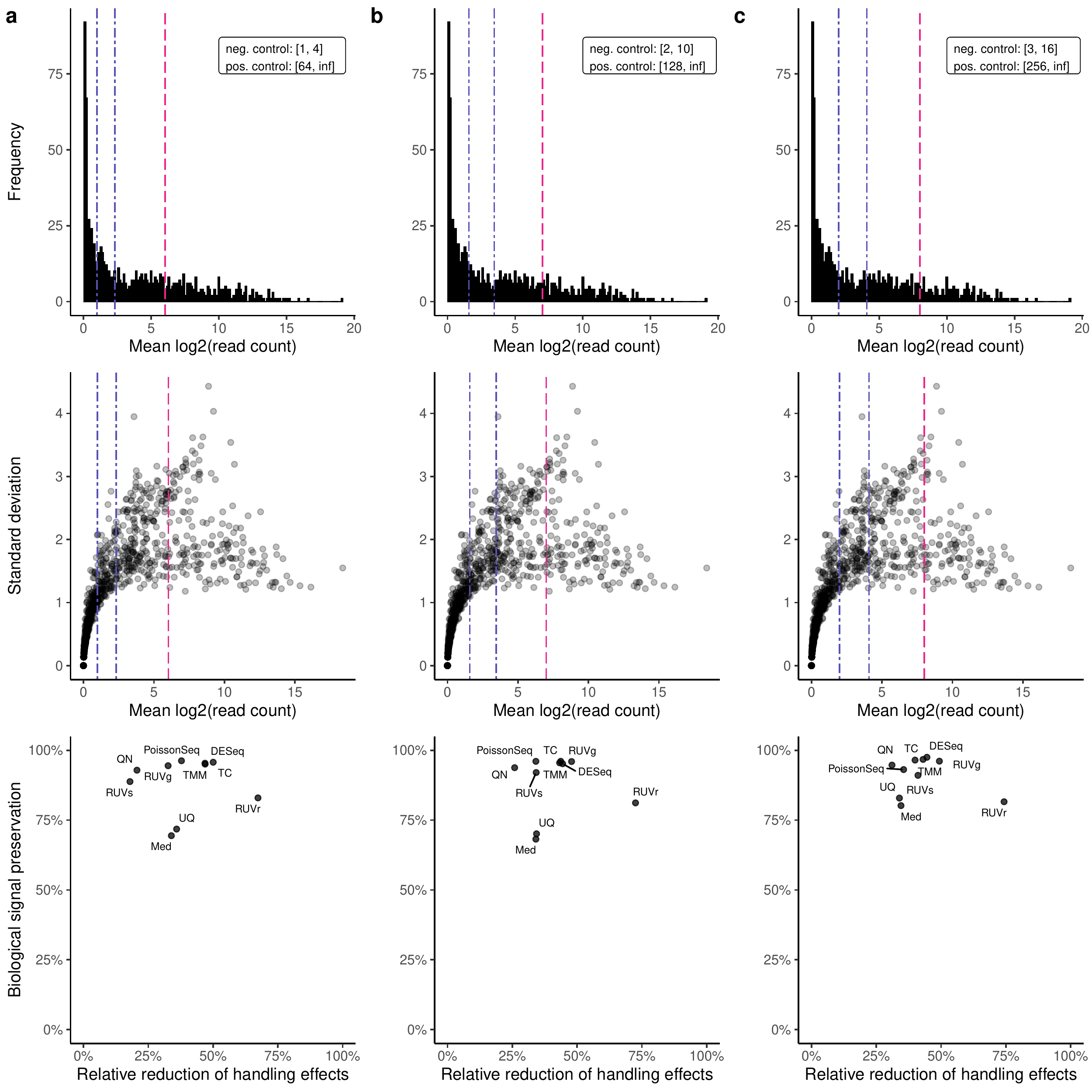} 
	\caption{
		Comparison of \method\ results for the MSK \testdata\ data between different choices of the ranges~$[\tzero,\tpoor]$ and~$[\twell,\infty)$ for negative and positive controls, respectively.
		The control definition is based on the mean count histograms (top) and the mean-standard deviation plots (middle).
		The \method\ result metrics (bottom) are shown for three different cutoff choices: configuration~\configa\ (left) with $[\tzero,\tpoor]=[1,4]$ and $[\twell,\infty)=[64,\infty)$; configuration~\configb\ (center) with $[\tzero,\tpoor]=[2,10]$ and $[\twell,\infty)=[128,\infty)$, which corresponds to the control definition in the main paper; and configuration~\configc\ (right) with $[\tzero,\tpoor]=[3,16]$ and $[\twell,\infty)=[256,\infty)$.
	}
	\label{fig:MSK-cutoff-study}
\end{figure}

We empirically study the sensitivity of the \method\ method for three different cutoff choices.
The three chosen configurations are \configa\ $[\tzero,\tpoor]=[1,4]$ and $[\twell,\infty)=[64,\infty)$; \configb\ $[\tzero,\tpoor]=[2,10]$ and $[\twell,\infty)=[128,\infty)$, which corresponds to the control definition in the main paper; and \configc\ $[\tzero,\tpoor]=[3,16]$ and $[\twell,\infty)=[256,\infty)$.
The number of negative controls are \configa: 92, \configb: 102, and \configc: 103, and the number of positive controls are \configa: 150, \configb: 115, and \configc: 77.
For each configuration, we followed the guidelines presented in the main paper and first selected \tzero\ such that selected poorly-expressed miRNAs show at least mild expression and then selected \tpoor and \twell such that each control group consists of a sufficient number of of markers (Fig.~\ref{fig:MSK-cutoff-study}, top and middle panel).

We observe that the cutoff choices only mildly affect the \method\ results (Fig.~\ref{fig:MSK-cutoff-study}, bottom).
For all configurations under study, we observe very similar relative \method\ metrics for all considered normalization methods; where, generally, DESeq and TMM offer the best trade-off between high reduction of handling effects and high preservation of biological signals.
Hence, we conclude that \method\ is not sensitive to the cutoff choices for defining the control markers. 

For all further results, we use the same definition of negative controls ($[\tzero,\tpoor]=[\msktesttzero,\msktesttpoor]$) and positive controls ($[\twell,\infty)=[\msktesttwell,\infty)$) as in the main paper.

\subsection{Method consideration and tuning parameter calibration for the estimation of partial correlations in positive controls}
\label{subsec:MSK-pcorr-estimation}

For positive controls, we examine different approaches for the estimation of partial correlations.
We compare the estimated partial correlation matrices and the \method\ results for the \testdata\ data using (1) the neighborhood selection by~\citet{Meinshausen.2006} (MB), (2) the glasso method \citep{Friedman.2008}, and (3) the FastGGM method~\citep{Ren2015}.
The MB method requires calibration of a tuning parameter, for which we consider Bayesian Information Criterion (BIC), Akaike Information Criterion (AIC), 10-fold Cross-Validation (CV), and Adaptive Validation (AV)~\citep{Chichignoud2016}.
We, further, consider tuning parameter calibration using the Stability Approach for Regularization Selection (StARS) and the Rotation Information Criterion (RIC) for MB and glasso, as implemented in the \texttt{R huge} package~\citep{Zhao2012}, and also the default tuning parameter calibration for the FastGGM method~\citep{Wang2016a}.

\begin{figure}
	\centering
	\includegraphics[width=0.99\textwidth]{\figpath/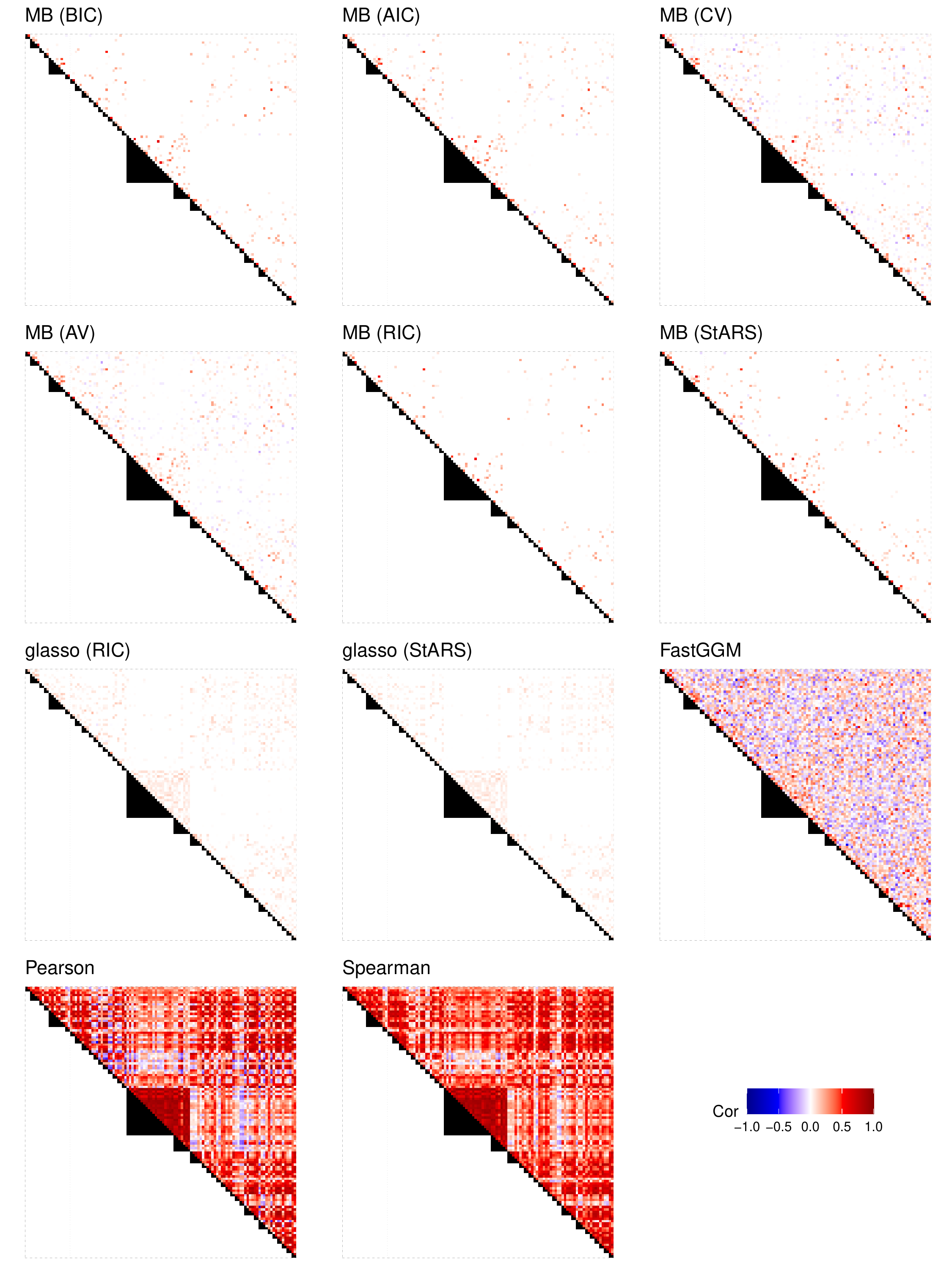} 
	\caption{
		Estimated correlations in positive controls for the un-normalized MSK \testdata\ data.
		The correlation matrix is displayed using heatmaps for each correlation estimation method under study.
		The three top rows show estimated partial correlations and the bottom row shows estimated marginal Pearson (bottom left) or Spearman (bottom center) correlations, respectively.
		In each heatmap, the upper triangular matrix shows the correlation strength and the lower triangular matrix indicates miRNA polycistronic clustering.
	}
	\label{fig:MSK-pcorr-all-pcorr-methods}
\end{figure}

We observe relatively similar correlation patterns for the un-normalized MSK \testdata\ data set using MB or glasso for all tuning parameter calibration methods (Fig.~\ref{fig:MSK-pcorr-all-pcorr-methods}).
The correlations estimated using MB, CV and AV show the most conservative tuning, while RIC shows the most liberal one with only few non-zero partial correlations.
Correlations estimated using glasso show similar patterns to MB; however, overall correlation strengths are all low in comparison to MB with only a few high partial correlations.
This indicates that glasso may be more suitable for graphical model estimation than partial correlation estimation and, hence, we do not use it as the defaul method for \method.
FastGGM estimated a dense partial correlation matrix, and, hence, may not be suitable for our application; this may be an artifact to its tuning parameter calibration and will be subject to future study.

Finally, we compare estimated partial correlations with Pearson and Spearman correlation.
In contrast to the partial correlation matrices estimated using MB or glasso, Pearson and Spearman correlation matrices estimate fully populated correlation matrices.
However, in positive controls, we intend to capture only the \textit{direct} co-expression relation for each pair of markers, removing any spurious correlation due to co-expression with other positive controls.
Hence, partial correlations are more suitable than marginal correlations.

\begin{figure}
	\centering
	\includegraphics[width=0.99\textwidth]{\figpath/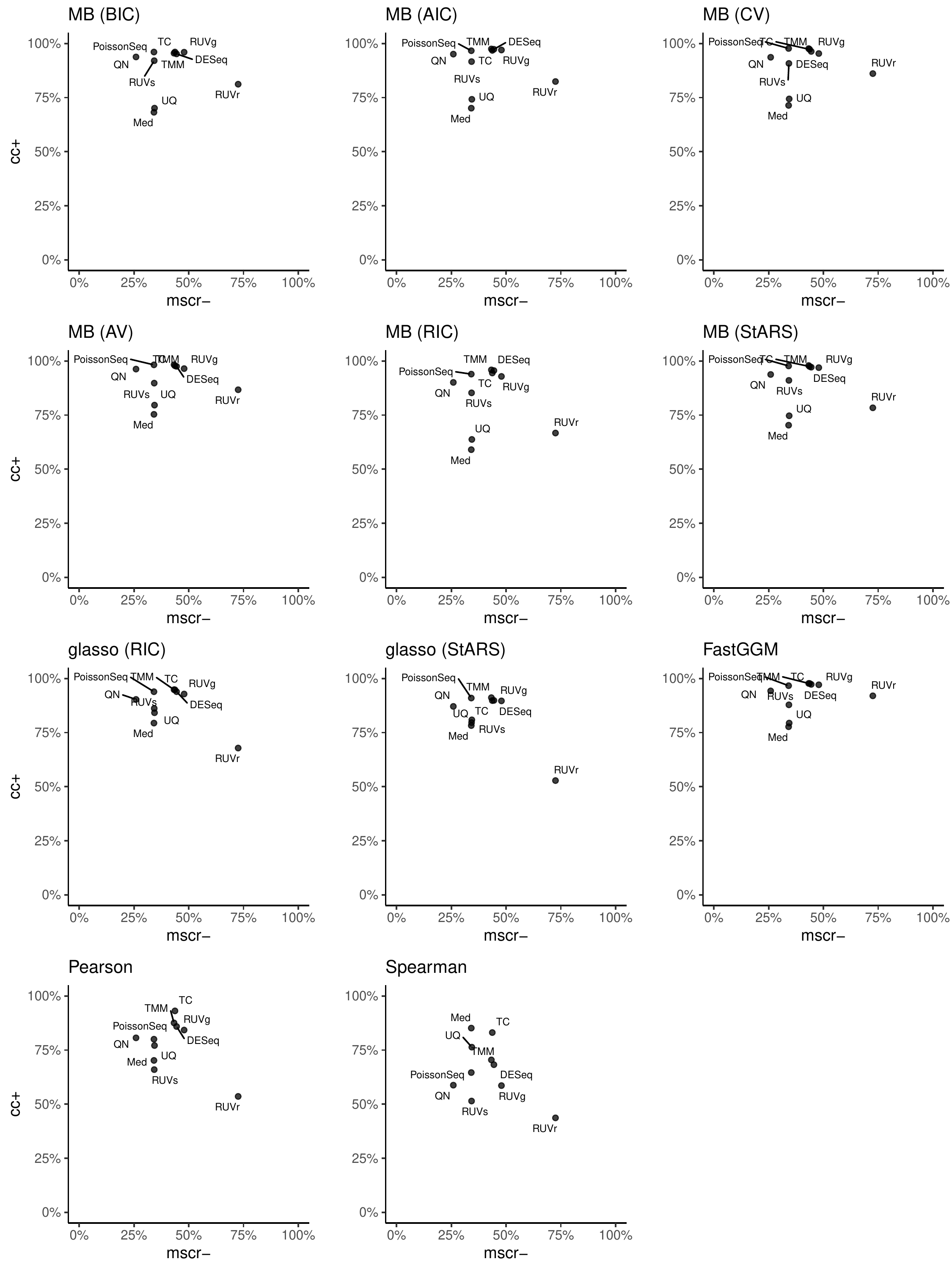} 
	\caption{
		Scatter plots of the two summary metrics for the MSK \testdata\ data using varying partial correlation estimation methods for positive controls.
		Each point stands for a normalization method under study.
		Note that varying the partial correlation estimation method, only the \cc\ metric changes and the \mscr\ metric, which is computed using negative controls only, stays constant across methods.
	}
	\label{fig:MSK-metrics-all-pcorr-methods}
\end{figure}

\begin{figure}
	\centering
	\includegraphics[width=0.99\textwidth]{\figpath/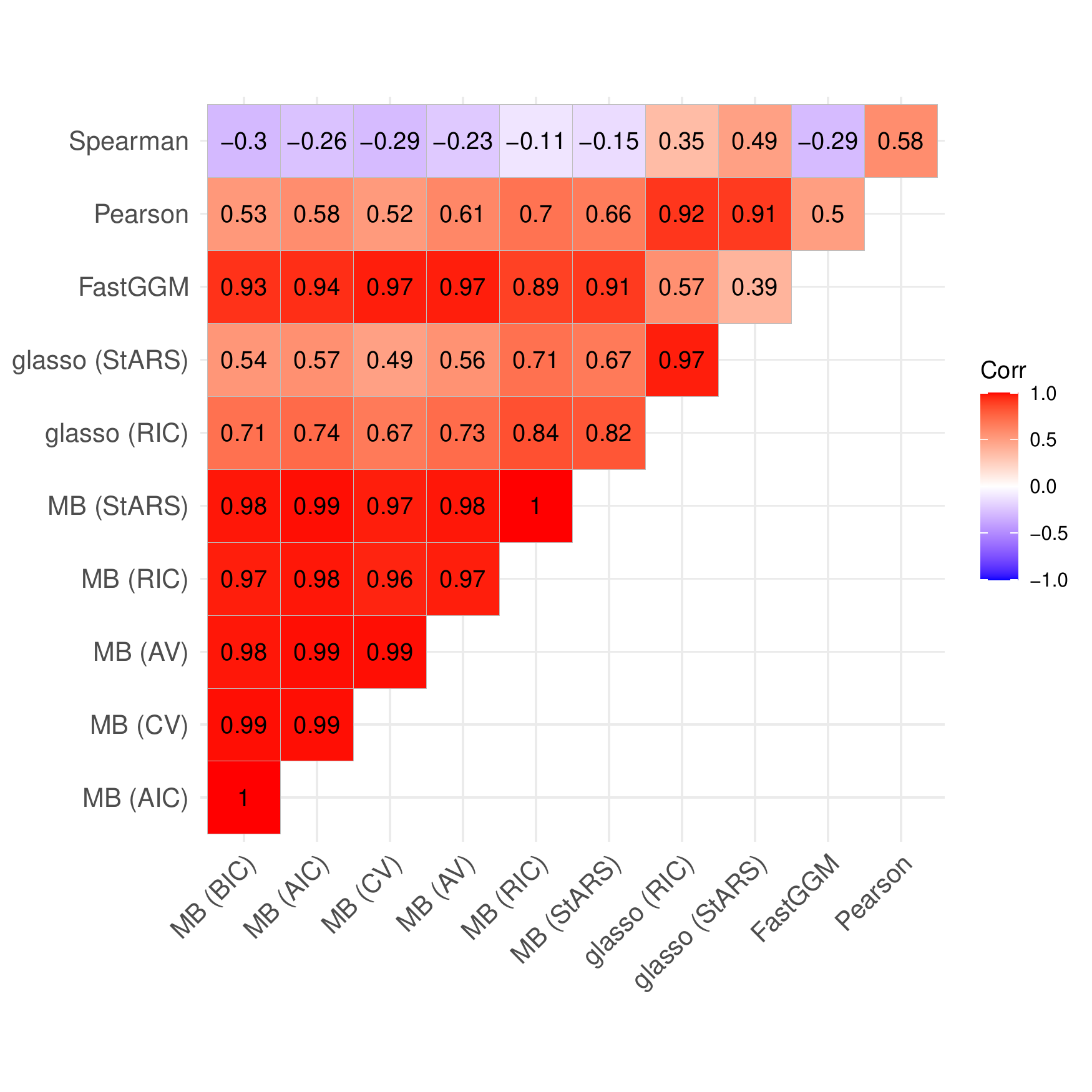} 
	\caption{
		Correlation of the \method\ \cc\ metric for positive controls across different partial correlation estimation methods.
	}
	\label{fig:MSK-pcorr-methods-corr}
\end{figure}

We apply our \method\ method to the MSK \testdata\ data using all of the aforementioned precision estimation methods (Fig.~\ref{fig:MSK-metrics-all-pcorr-methods}).
We observe that using the MB method the relative performance of each normalization method stays the same regardless of the tuning parameter calibration used for MB.
Similarly, comparing the relative \cc\ metrics of normalization methods, glasso yields similar results to MB for both RIC and StARS.
Indeed, we observe very high correlation of the \method\ \cc\ metrics across all MB methods, high correlation of MB with glasso+RIC, but only moderate correlation between MB and glasso+StARS  (Fig.~\ref{fig:MSK-pcorr-methods-corr}).
Even though FastGGM estimated a dense correlation matrix, we observe high to very high correlation with MB depending on the tuning parameter calibration used for MB.

We reason that it is essential for \method\ to accurately estimate the relative strengths of within-cluster partial correlations, which are used for computing the \cc\ metric, rather than correctly estimating the sparsity of partial correlations;
hence, MB and FastGGM are more suitable for the \method\ method than glasso. 
We decide to use the MB method as a default for the \method\ method since it correctly estimates a sparse partial correlation matrix.
We use BIC tuning as its computation is fast and simple.

\subsection{Partial correlations of positive controls in the \testdata\ data}
\label{subsec:MSK-pcorr-plots}

\begin{figure}
	\centering
	\includegraphics[width=0.99\textwidth]{\figpath/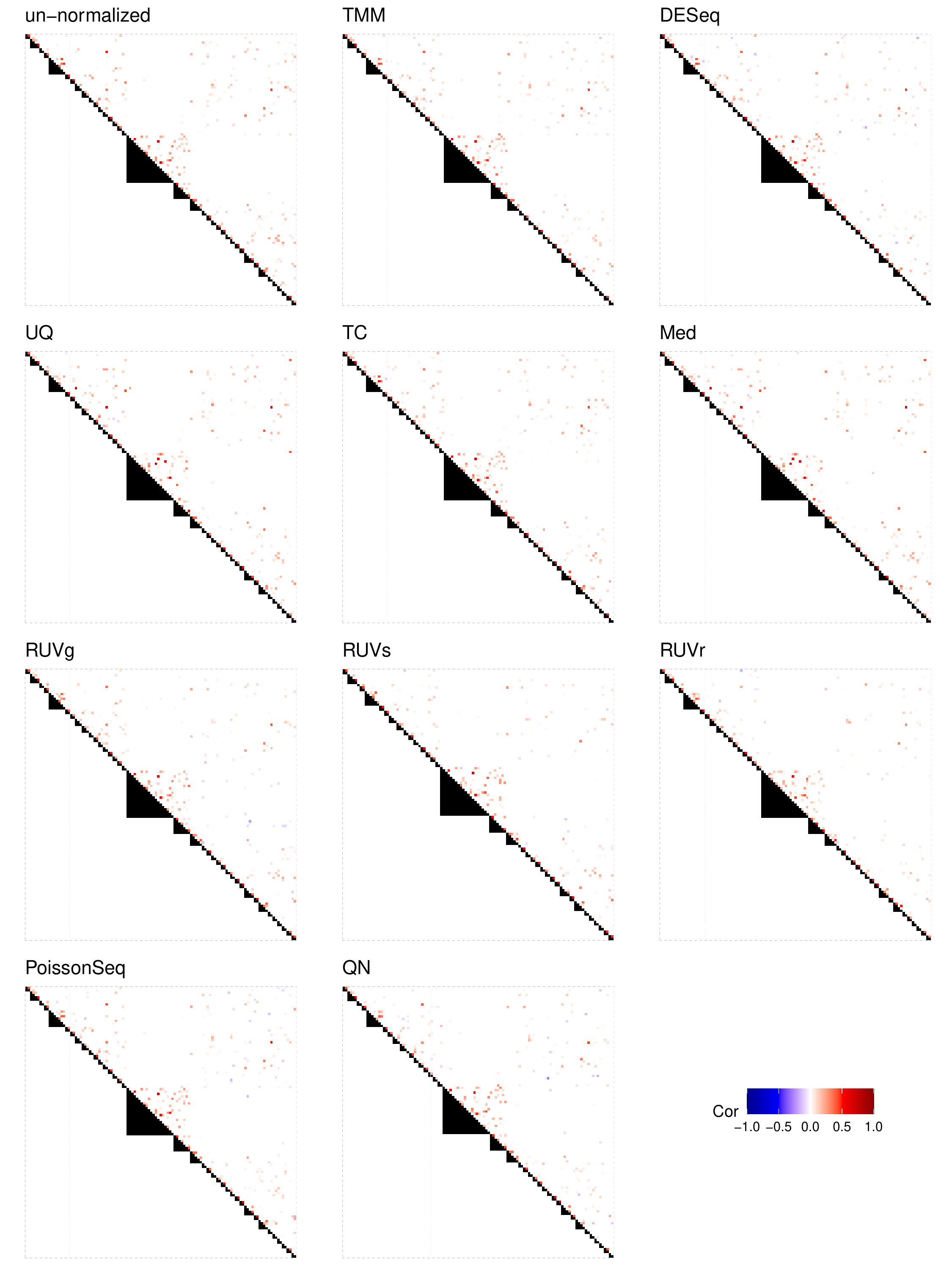} 
	\caption{
		Heatmaps of the estimated correlations among positive controls in the normalized MSK \testdata\ data for each normalization method and for un-normalized MSK \testdata\ data.
		In each heatmap, the upper triangular matrix shows the correlation strength and the lower triangular matrix indicates miRNA polycistronic clustering.
	}
	\label{fig:MSK-pcorr-all-norm-methods}
\end{figure}

\begin{figure}
	\centering
	\includegraphics[width=0.99\textwidth]{\figpath/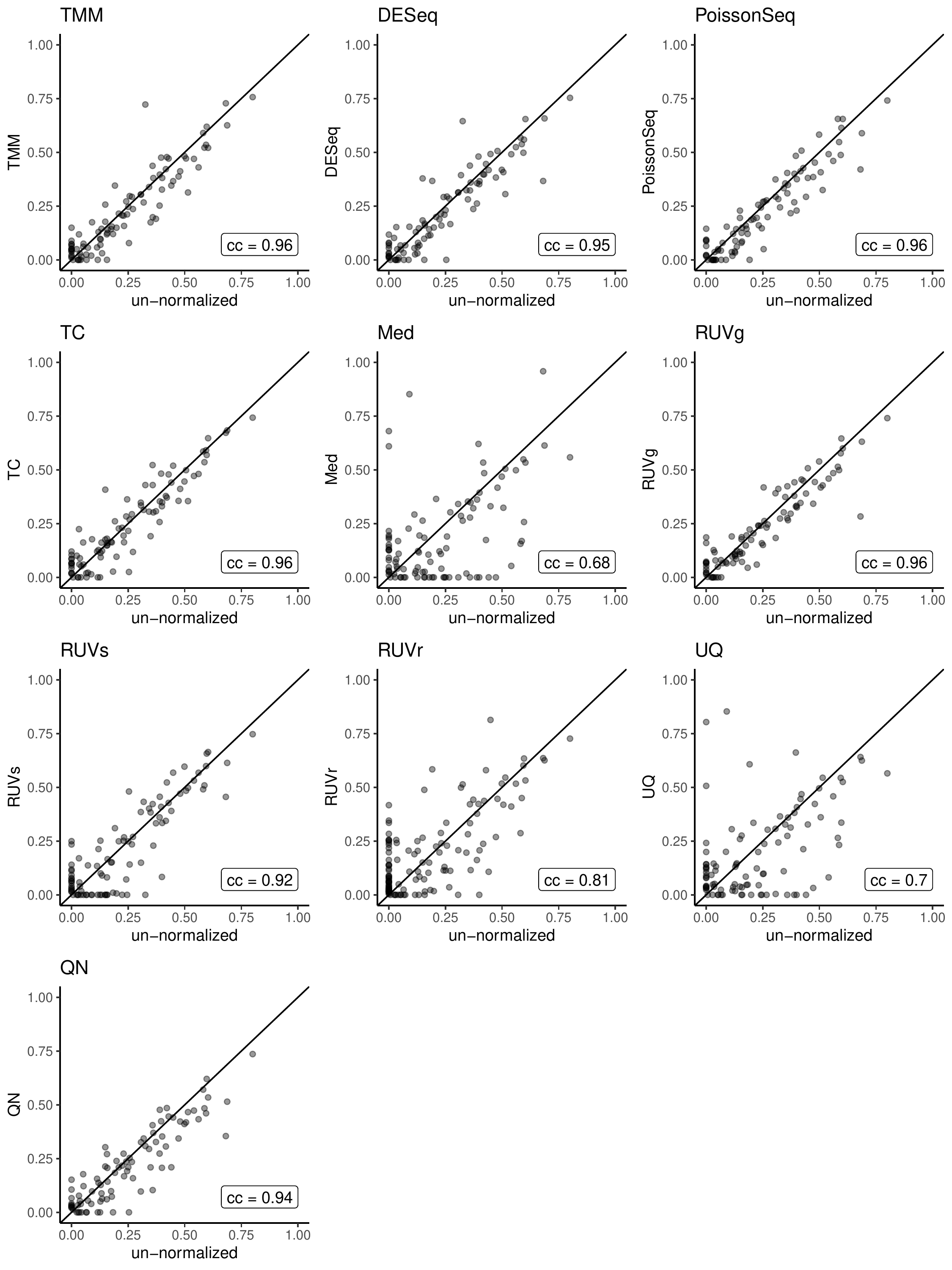} 
	\caption{
		Scatter plots of partial correlations in positive controls before versus after normalization for each normalization method in the MSK \testdata\ data.
		The corresponding concordance correlation coefficient between the shown partial correlations, which is measured by~\cc, is indicated in the lower right corner of each plot.
	}
	\label{fig:MSK-pcorr-scatter-all}
\end{figure}

For positive controls, all within-cluster partial correlations in the \testdata\ data are strictly positive as expected regardless of normalization (Fig.~\ref{fig:MSK-pcorr-all-norm-methods}).
Furthermore we observe a high abundance of strong positive partial correlations for clustered markers as we expected based on biological evidence reported in the literature on polycistronic clusters.
Compared to the \benchmarkdata\ data (see the main paper), the \testdata\ data has fewer positive within-cluster correlations and more positive, off-cluster correlations, both of which likely resulted from excessive handling effects.
Normalization, regardless of the method, alleviated the off-cluster correlations in terms of both the number and strength; 
depending on the method, normalization can either retain or reduce within-cluster correlations, which is signified by their before-versus-after-normalization concordance. 
This concordance strongly varies with the used normalization method (Fig.~\ref{fig:MSK-pcorr-scatter-all}). TMM ($\cc=0.96)$, DESeq ($0.95$), PoissonSeq ($0.96$), TC ($0.96$), RUVg ($0.96$), and RUVs ($0.92$) all show very high concordance as measured by the metric \cc.
RUVr ($0.81$), UQ ($0.70$), and Med ($0.68$) fail to achieve such high \cc\ with a much bigger spread of corresponding correlations in the partial correlation scatter plots compared to the aforementioned methods.

\subsection{Marginal correlations of negative controls in the \benchmarkdata\ and \testdata\ data}
\label{subsec:MSK-corr-freq-curves}

\begin{figure}
	\centering
	\includegraphics[width=0.99\textwidth]{\figpath/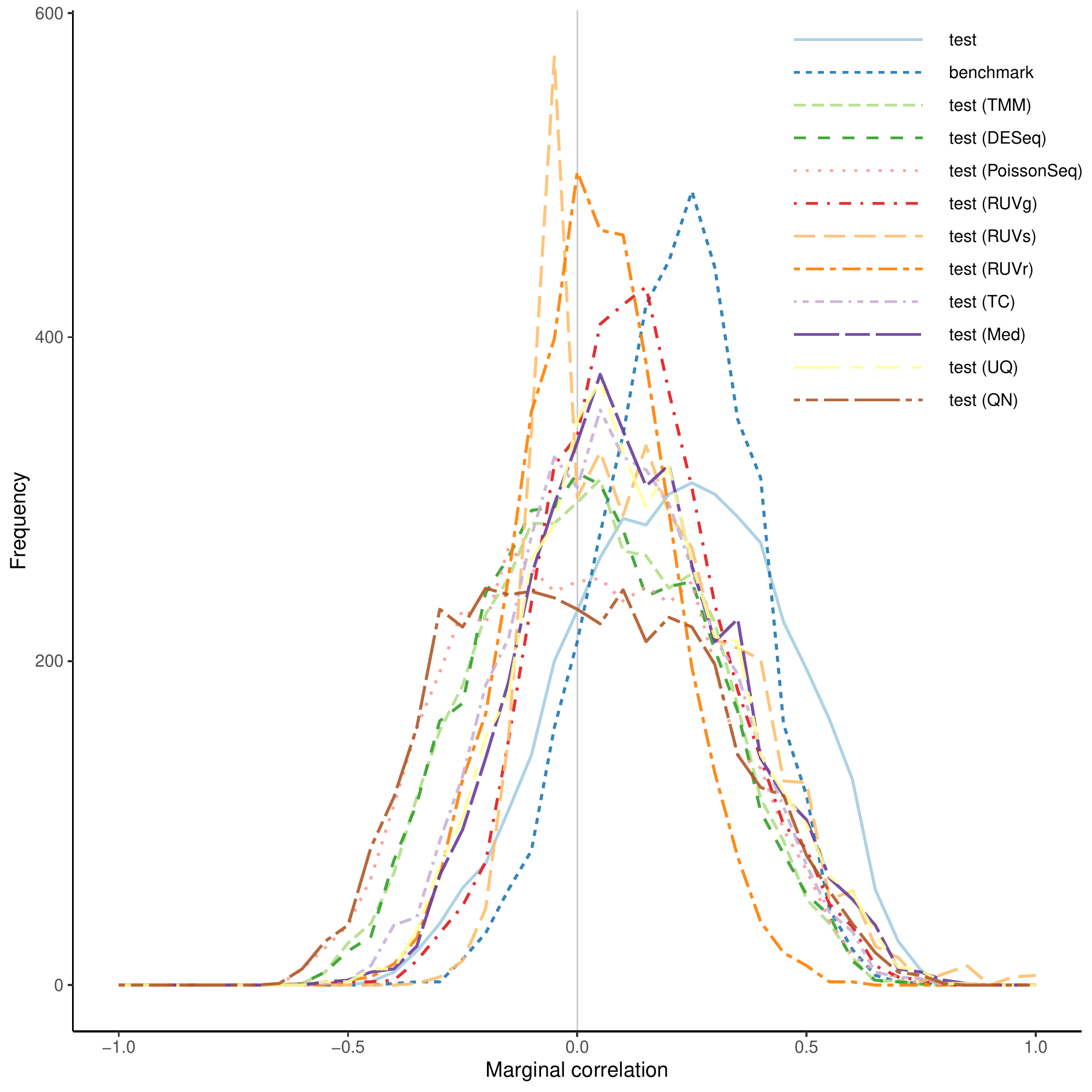} 
	\caption{
		Histogram frequency curve for marginal correlations in negative controls for the un-normalized \benchmarkdata\ data, un-normalized \testdata\ data, and \testdata\ data normalized using each normalization method under study.
	}
	\label{fig:MSK-corr-freq-curves}
\end{figure}

For negative controls, we assess the level of inter-marker correlations in the \testdata\ data before and after normalization, and compare these with that in the benchmark data (Fig.~\ref{fig:MSK-corr-freq-curves}).
The frequency of strong correlations in the \testdata\ data is much higher compared to the \benchmarkdata\ data while, generally, most correlations in the \benchmarkdata\ data are weak (centered around 0.21).
The variation of correlation strengths in the \testdata\ data is much higher, and, in particular, stronger positive correlations are much more abundant compared to the \benchmarkdata\ data.
This finding accords with our assumption that handling effects manifest themselves as excessively strong positive correlations.
The level of correlation strengths and correlation variability after normalization vary among the different normalization methods with some methods leading to a stronger correlation reduction compared to others;
here, RUVr stands out from the other methods and achieved the highest correlation reduction. 
Furthermore, note that the correlation histogram curves for TC, UQ, and Med are very similar due to the similarity of these methods.

\begin{figure}
	\centering
	\includegraphics[width=0.99\textwidth]{\figpath/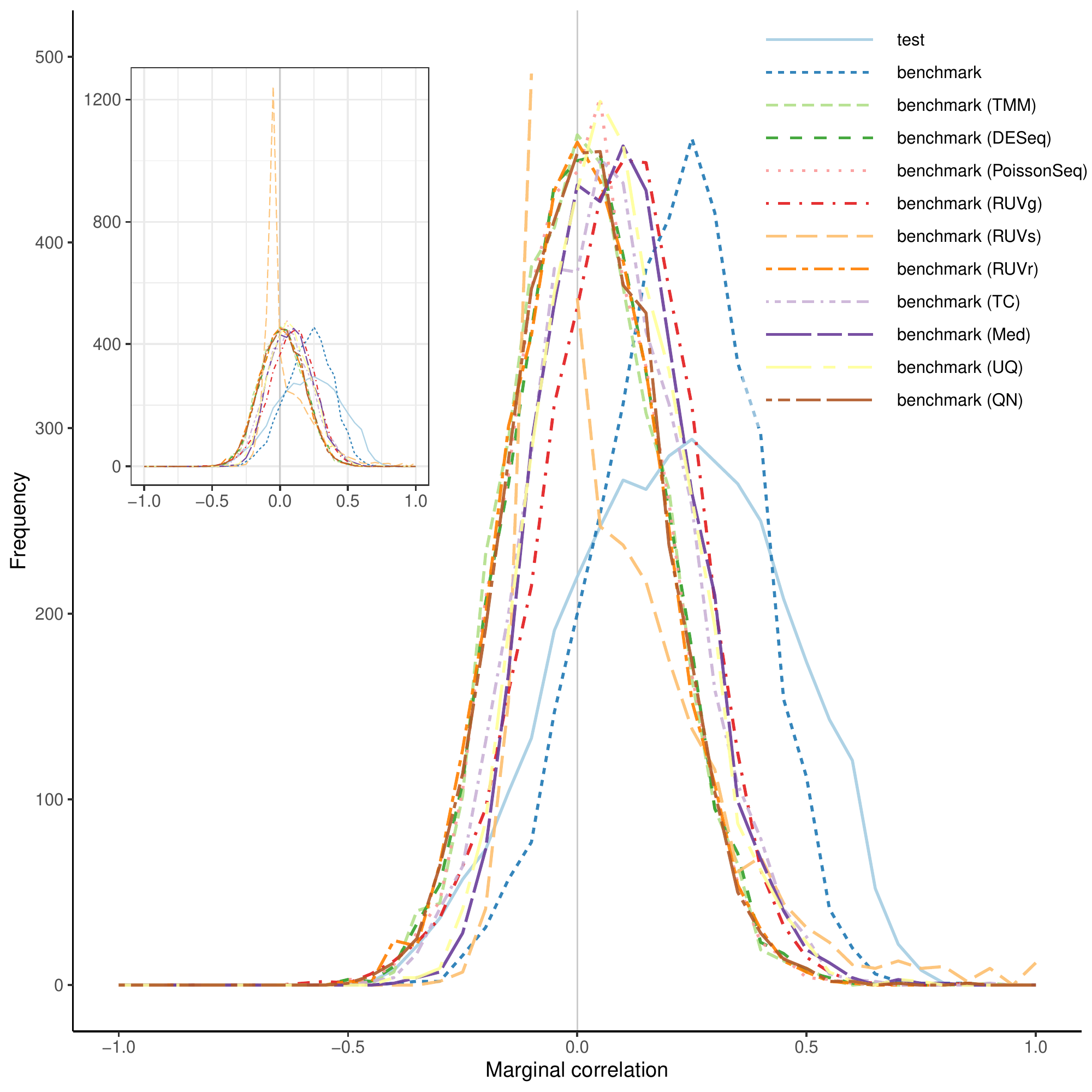} 
	\caption{
		Histogram frequency curve for marginal correlations in negative controls for the un-normalized \benchmarkdata\ data, un-normalized \testdata\ data, and \benchmarkdata\ data normalized using each normalization method under study.
	}
	\label{fig:MSK-corr-freq-curves-benchmark}
\end{figure}

For normalized \benchmarkdata\ data (Fig.~\ref{fig:MSK-corr-freq-curves-benchmark}), we observe that most normalization methods show very similar correlation variance.
However, some methods (RUVg, TMM, DESeq, and QN) show better centering than the others. 
RUVs stands out, since it introduces some strong positive correlations and shows an overall unbalanced correlation pattern with a spike for small negative correlations.
Hence, RUVs is not suitable for normalization of the \benchmarkdata\ data.

\clearpage
\section{TCGA-UCEC Data}
\label{sec:tcga-ucec-data}

\subsection{Data Description} 
\label{subsec:UCEC-data-description}

We constructed the two data sets so that they each contain \ucecnumsamples~samples, \ucecnumsamplesEND\ of which were of endometrioid subtype (END) and \ucecnumsamplesSER\ of serous subtype (SER), and so that these two data sets have 24 samples in common.
The \ucecnumsamples~samples of the first data set were processed in a single batch (batch number 228.63.0), which we refer to as the ``\singlebatchdata'' data.
The second data set (that is the ``\mixedbatchdata'') is composed of 24~samples (11~END and 13~SER) from the first data set (batch 228.63.0) and 24~samples (11~END and 13~SER) from other batches; END samples are from the batches 104.85.0~(1 sample), 121.83.0~(1), 156.80.0~(1), 186.74.0~(2), 59.83.0~(2), 73.85.0~(2), 81.83.0~(1), and 92.87.0~(1); and SER samples are from the batches 110.75.0~(1), 137.81.0~(1), 143.82.0~(1), 156.80.0~(1), 168.78.0~(2), 178.78.0~(1), 186.74.0~(1), 201.70.0~(1), 324.54.0~(1), 381.43.0~(1), 49.86.0~(1), and 73.85.0~(1).
We refer to the latter data set as the ``\mixedbatchdata'' data.
Sequencing counts are available for \ucecnumgenes~miRNAs in each sample.

\subsection{Correlation in positive and negative controls}
\label{subsec:UCEC-corr}

\begin{figure}
	\centering
	\includegraphics[width=0.99\textwidth]{\figpath/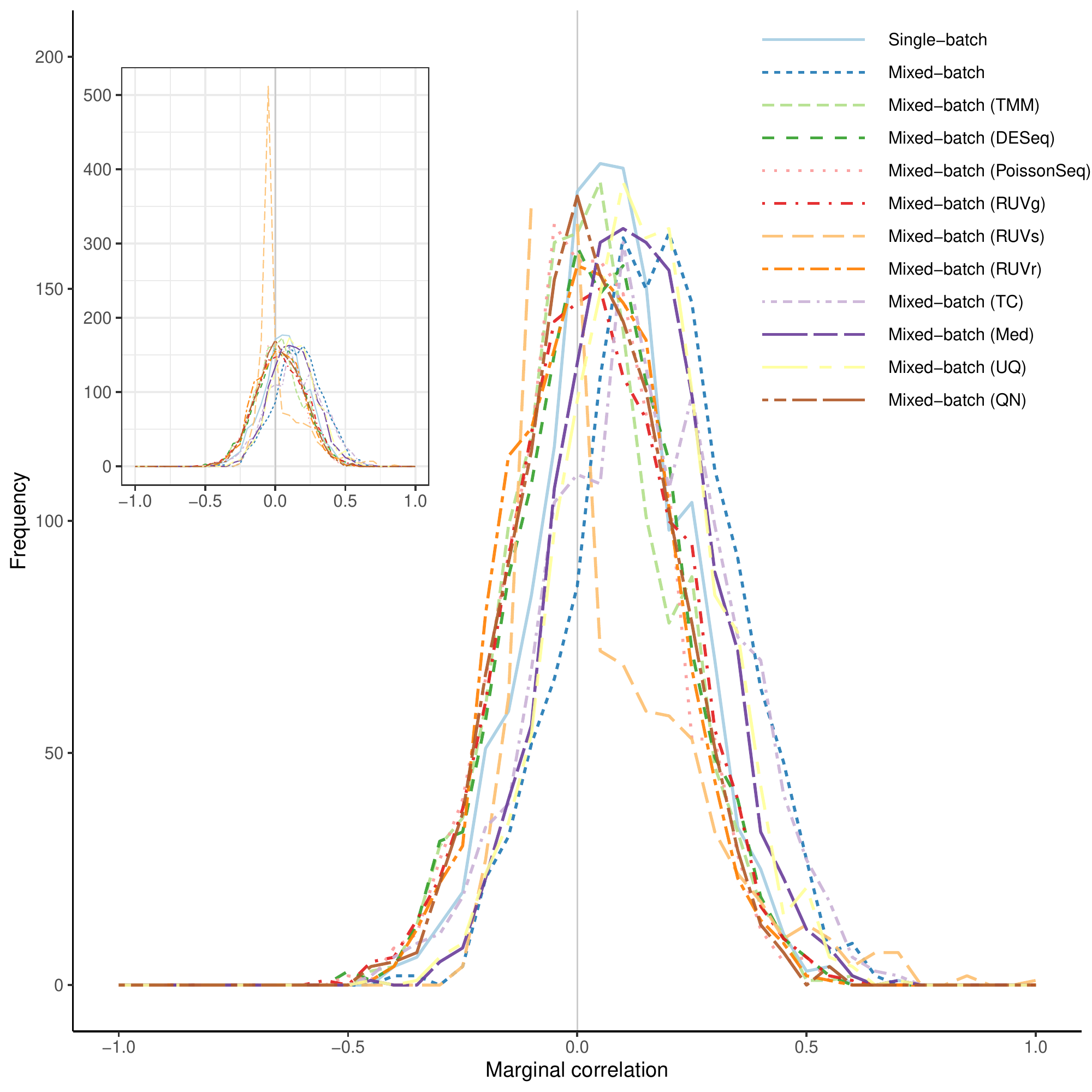} 
	\caption{
		Histogram frequency curve for marginal correlations in negative controls for the un-normalized \singlebatchdata\ data, un-normalized \mixedbatchdata\ data, and normalized \mixedbatchdata\ data using each normalization method under study.
	}
	\label{fig:UCEC-corr-freq-curves}
\end{figure}

For negative controls, we assess the level of their inter-marker Pearson correlation in the \mixedbatchdata\ data before and after normalization, and compare these with that in the \singlebatchdata\ data (Fig.~\ref{fig:UCEC-corr-freq-curves}). 
As for the MSK data sets, the frequency of strong correlations in the \mixedbatchdata\ data is higher compared to the \singlebatchdata\ data while, generally, most correlations in the \singlebatchdata\ data are weak.
This finding, again, accords with our assumption that handling effects manifest themselves as excessively strong positive correlations.
Normalization of the \mixedbatchdata\ data not only drastically reduces strong positive correlations but also shifts positive correlations towards zero for most normalization methods;
except for TC, UQ, Med, and RUVs, positive correlations in negative controls are alleviated below the level of correlations in the \singlebatchdata\ data.
Hence, these methods provide effective removal of handling effects in the handling-prone \mixedbatchdata\ data.

\begin{figure}
	\centering
	\includegraphics[width=0.99\textwidth]{\figpath/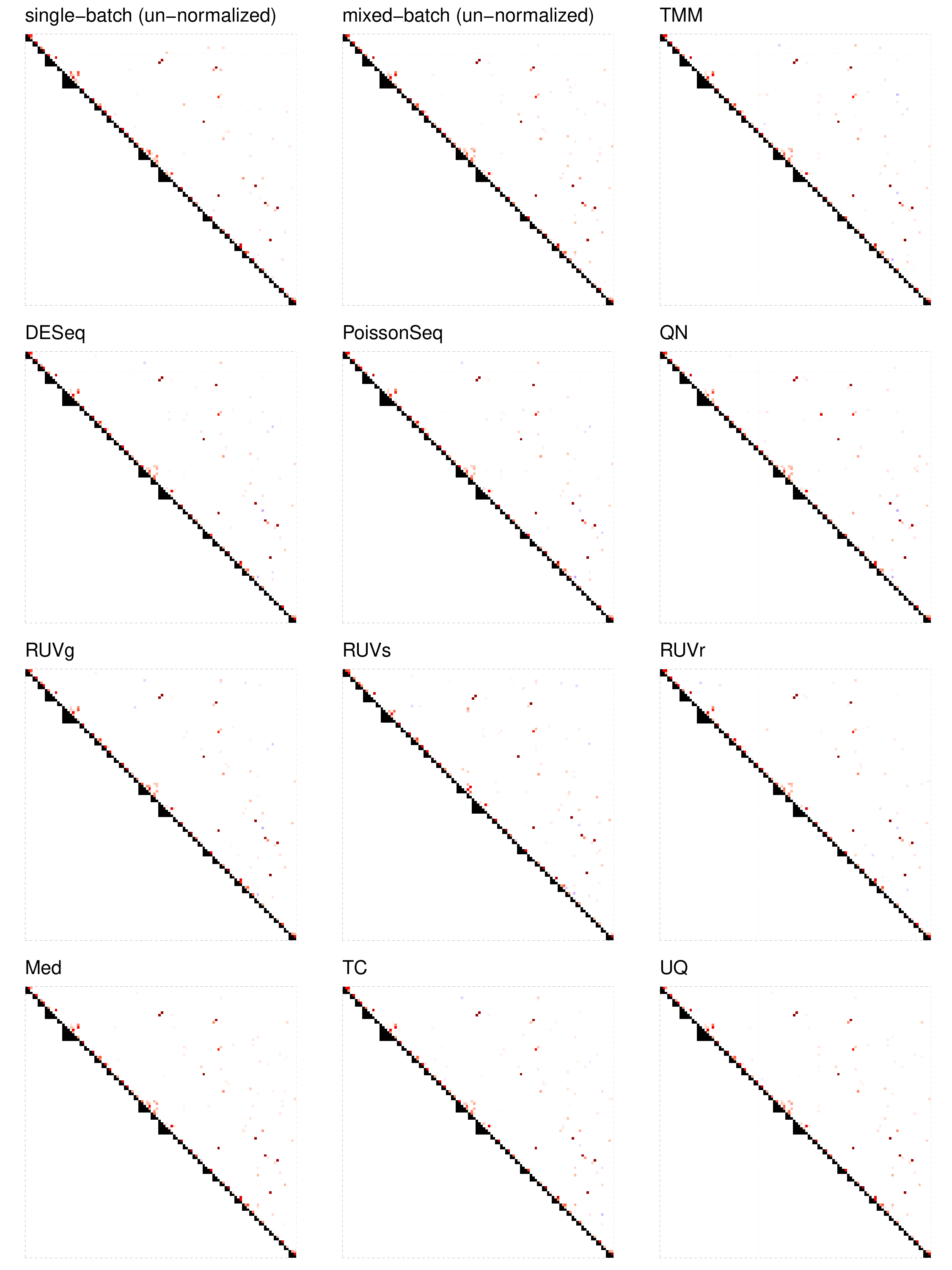} 
	\caption{
		Heatmaps of the estimated correlations among positive controls in the normalized TCGA-UCEC \mixedbatchdata\ data for each normalization method as well as for un-normalized TCGA-UCEC \singlebatchdata\ (top left) and un-normalized \mixedbatchdata\ (top center) data.
		In each heatmap, the upper triangular matrix shows the correlation strength and the lower triangular matrix indicates miRNA polycistronic clustering.
	}
	\label{fig:UCEC-pcorr-all-norm-methods}
\end{figure}

\begin{figure}
	\centering
	\includegraphics[width=0.99\textwidth]{\figpath/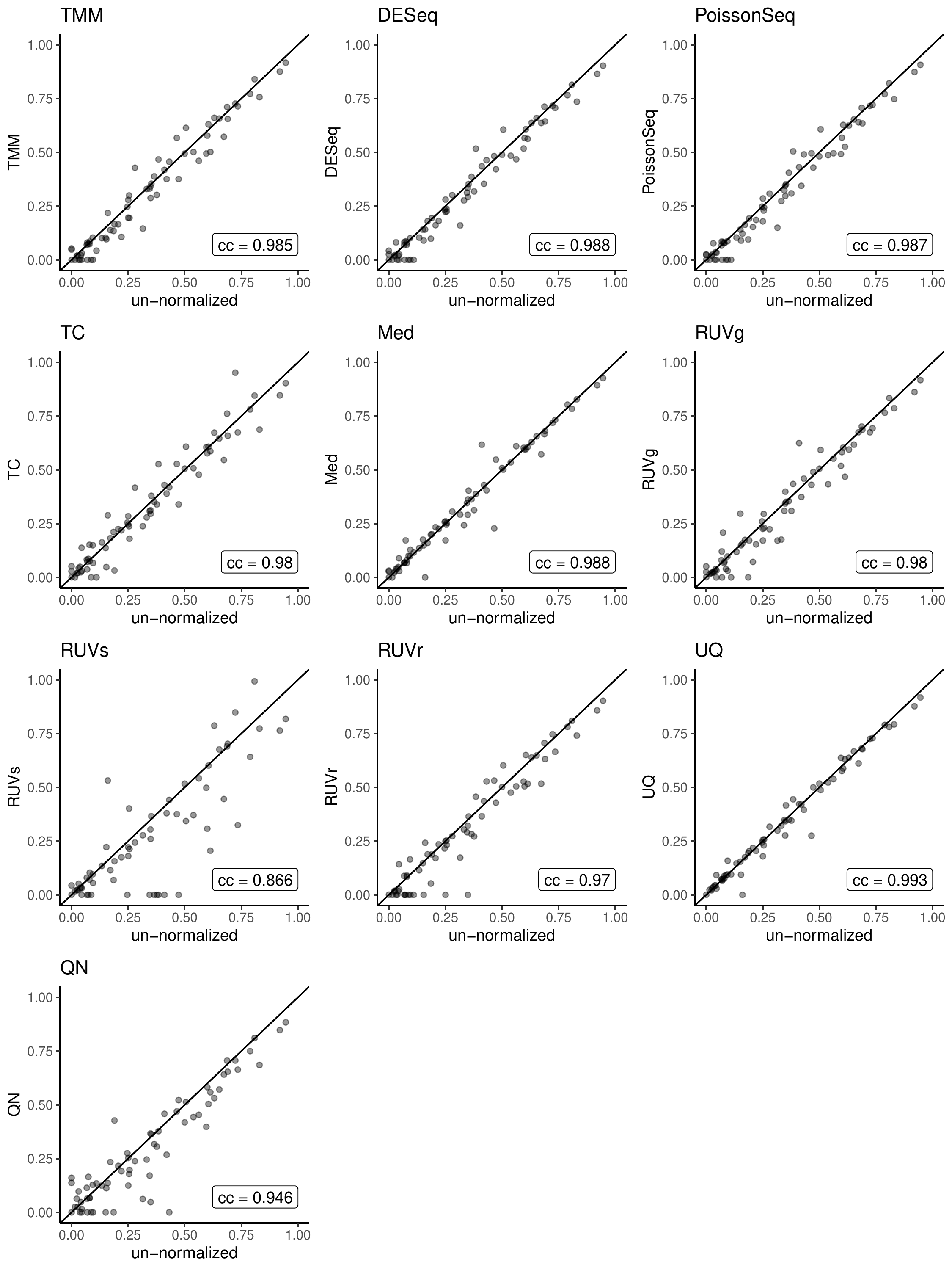} 
	\caption{
		Scatter plots of partial correlations in positive controls before versus after normalization for each normalization method for the TCGA-UCEC \mixedbatchdata\ data.
		The corresponding concordance correlation coefficient between the shown partial correlations, which is measured by~\cc, is indicated in the lower right corner of each plot.
	}
	\label{fig:UCEC-pcorr-scatter-all}
\end{figure}

For positive controls, we observe a high abundance of within-cluster correlations in the \singlebatchdata\, all of which are strictly positive as expected (Fig.~\ref{fig:UCEC-pcorr-all-norm-methods}).
As we have reported in the comparison of the MSK \benchmarkdata\ and \testdata\ data, the \mixedbatchdata\ data has fewer positive within-cluster correlations and more off-cluster positive correlations compared to the \singlebatchdata\ data.
Both likely, again, resulted from the added handling effects in the \mixedbatchdata\ data.
Normalization, regardless of the method, alleviated the off-cluster correlations in terms of both the number and strength.
Finally, depending on the method, normalization can either retain or eliminate within-cluster correlations, which is signified by their before-versus-after-normalization concordance, that is, the \cc\ metric (Fig.~\ref{fig:UCEC-pcorr-scatter-all}).

\clearpage
\section{TCGA-BRCA and TCGA-UCS Data}
\label{sec:brca-ucs-data}

\subsection{Control marker definition}
\label{subsec:BRCA-UCS-cutoff-study}

\begin{figure}
	\centering
	\includegraphics[width=0.49\textwidth]{\figpath/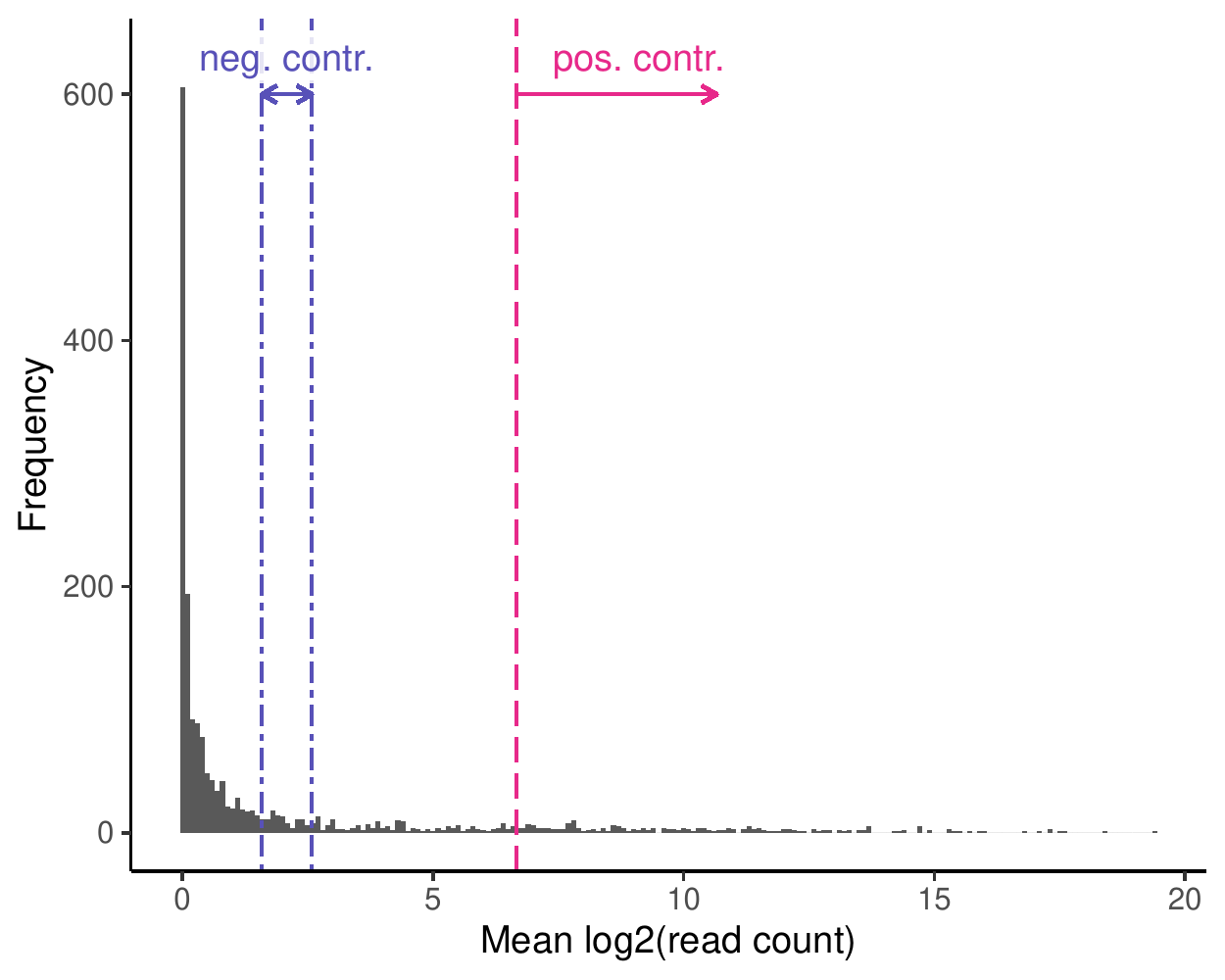} 
	\centering
	\includegraphics[width=0.49\textwidth]{\figpath/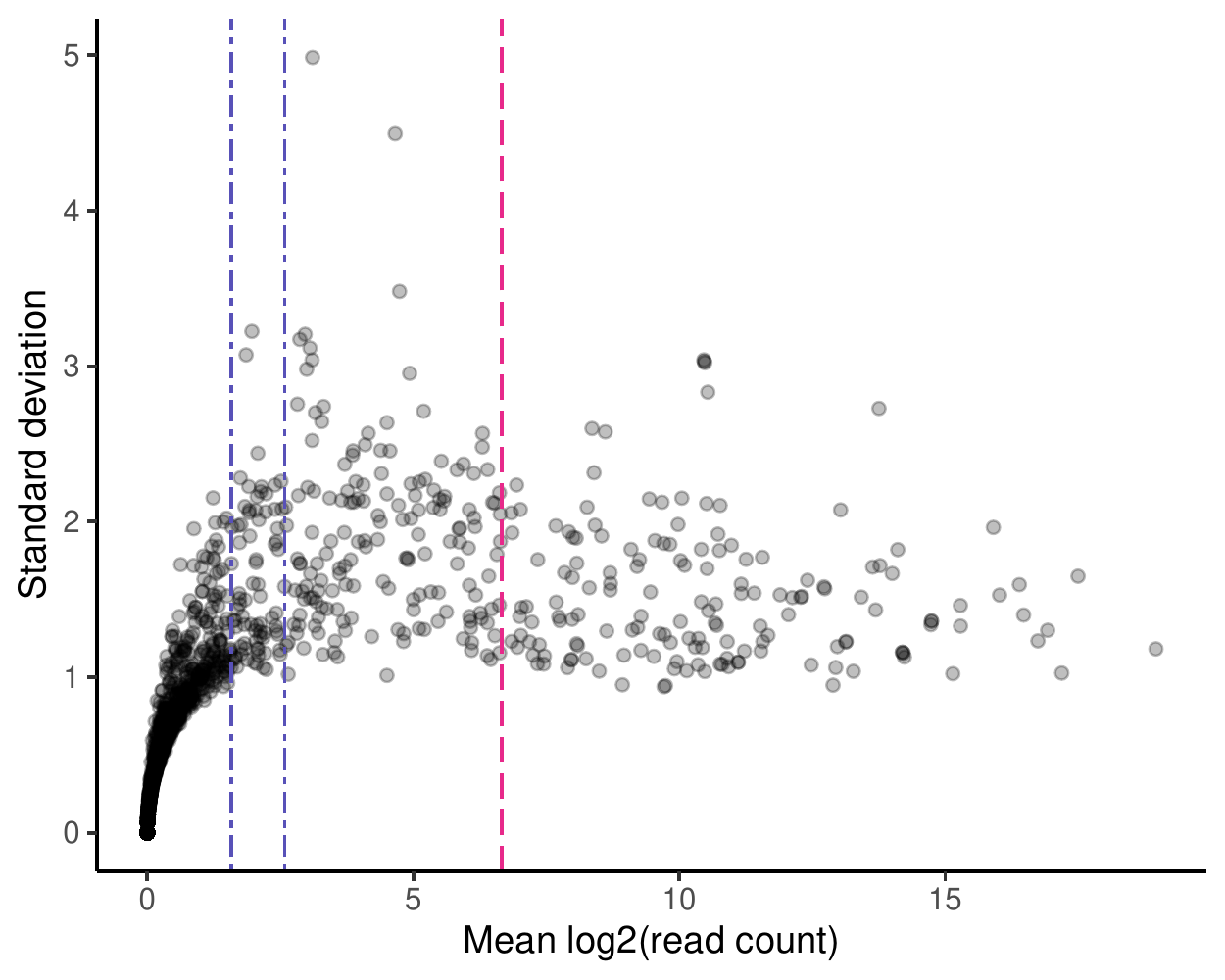} 
	\caption{
		Mean count histogram (left) and mean-standard deviation plot (right) for the TCGA-BRCA/UCS data.
		The ranges~$[\tzero,\tpoor]=[\tcgatzero,\tcgatpoor]$ for negative controls and~$[\twell,\infty)=[\tcgatwell,\infty)$ for positive controls are indicated by blue and red vertical lines, respectively.
	}
	\label{fig:BRCA-UCS-control-choice}
\end{figure}

Recall that the TCGA-BRCA/TCGA-UCS data set combines \tcganumsamplesBRCA~stage III and IV samples from the TCGA-BRCA project with all \tcganumsamplesUCS~samples from the TCGA-UCS project.
This data set does not have an available benchmark.
We selected the ranges~$[\tzero,\tpoor]=[\tcgatzero,\tcgatpoor]$ and~$[\twell,\infty)=[\tcgatwell,\infty)$ based on our recommended data-driven graphical criteria (Fig.~\ref{fig:BRCA-UCS-control-choice}). 
The number of negative controls is~$\numneg=\tcganumneg$ and the number of positive controls is~$\numpos=\tcganumpos$.

\subsection{Correlation in positive and negative controls}
\label{subsec:BRCA-UCS-corr}

\begin{figure}
	\centering
	\includegraphics[width=0.99\textwidth]{\figpath/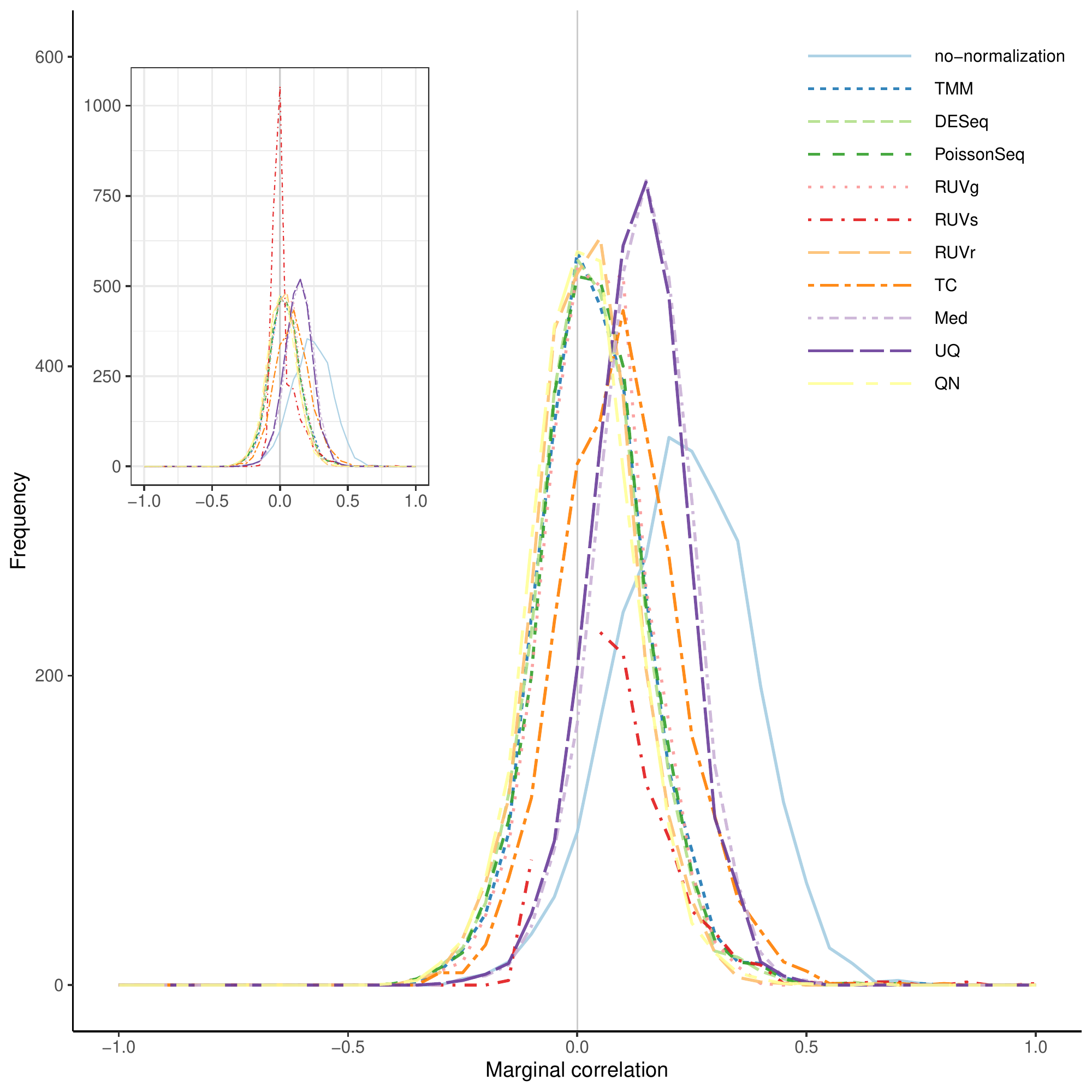} 
	\caption{
		Histogram frequency curve for marginal correlations in negative controls for the un-normalized TCGA-BRCA/UCS data and normalized TCGA-BRCA/UCS data for each normalization method under study.
	}
	\label{fig:BRCA-UCS-corr-freq-curves}
\end{figure}

For negative controls, we assess the level of their inter-marker correlation in the combined TCGA-BRCA and TCGA-UCS data set (Fig.~\ref{fig:BRCA-UCS-corr-freq-curves}). 
As for the MSK and TCGA-UCEC data sets, normalization drastically reduces strong positive correlations in negative controls and different normalization leads to different reduction of correlations and, hence, different reduction of handling effects, which is quantified by the \mscr\ metric.
Note further that similar normalization methods lead to similar correlation distributions; for example, Med and UQ normalization show a near-identical distribution.

\begin{figure}
	\centering
	\includegraphics[width=0.99\textwidth]{\figpath/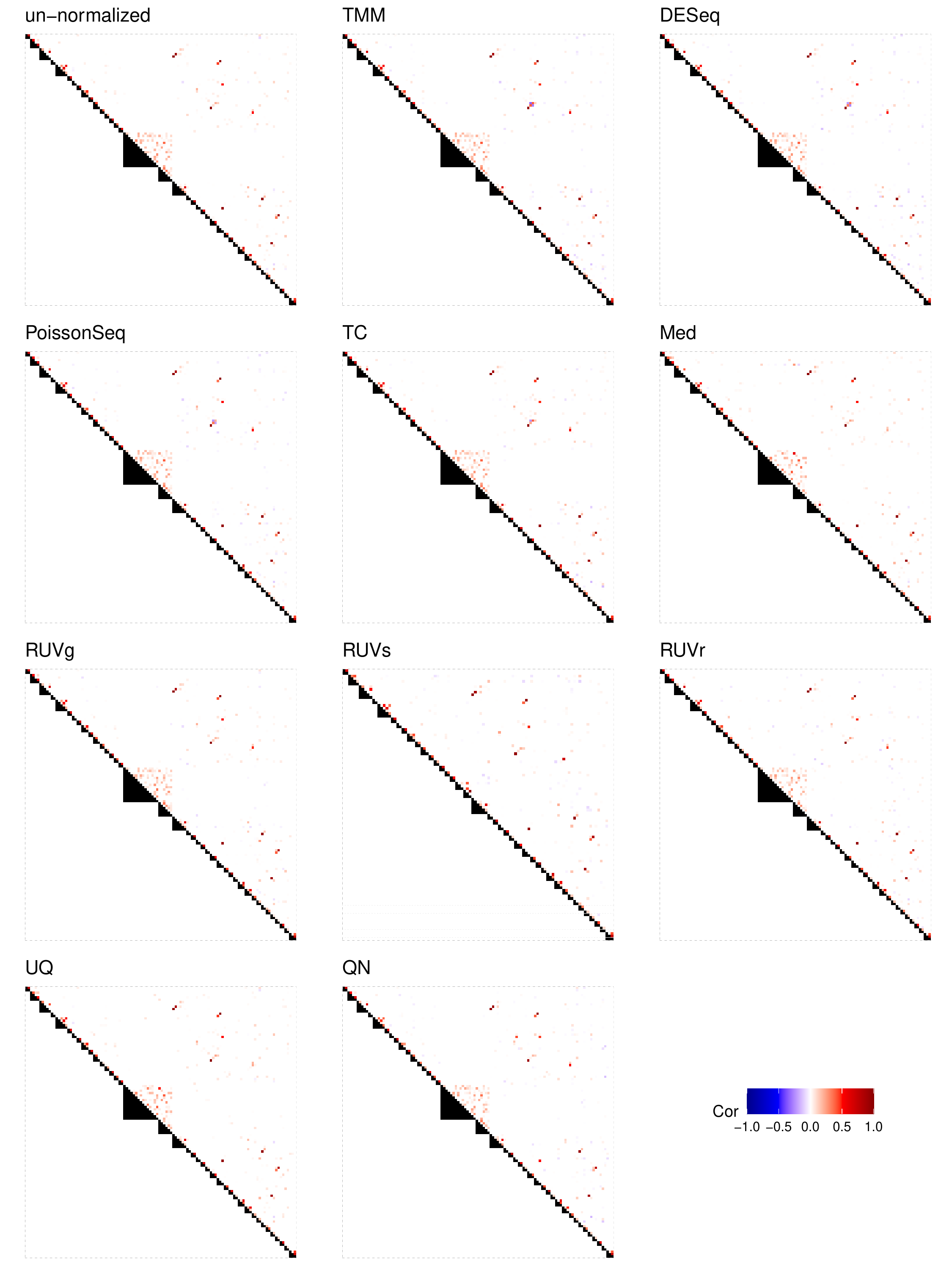} 
	\caption{
		Heatmaps of the estimated correlations among positive controls of the TCGA-BRCA/UCS data set before and after normalization.
		In each heatmap, the upper triangular matrix shows the correlation strength and the lower triangular matrix indicates miRNA polycistronic clustering.
	}
	\label{fig:BRCA-UCS-pcorr-all-norm-methods}
\end{figure}

\begin{figure}
	\centering
	\includegraphics[width=0.99\textwidth]{\figpath/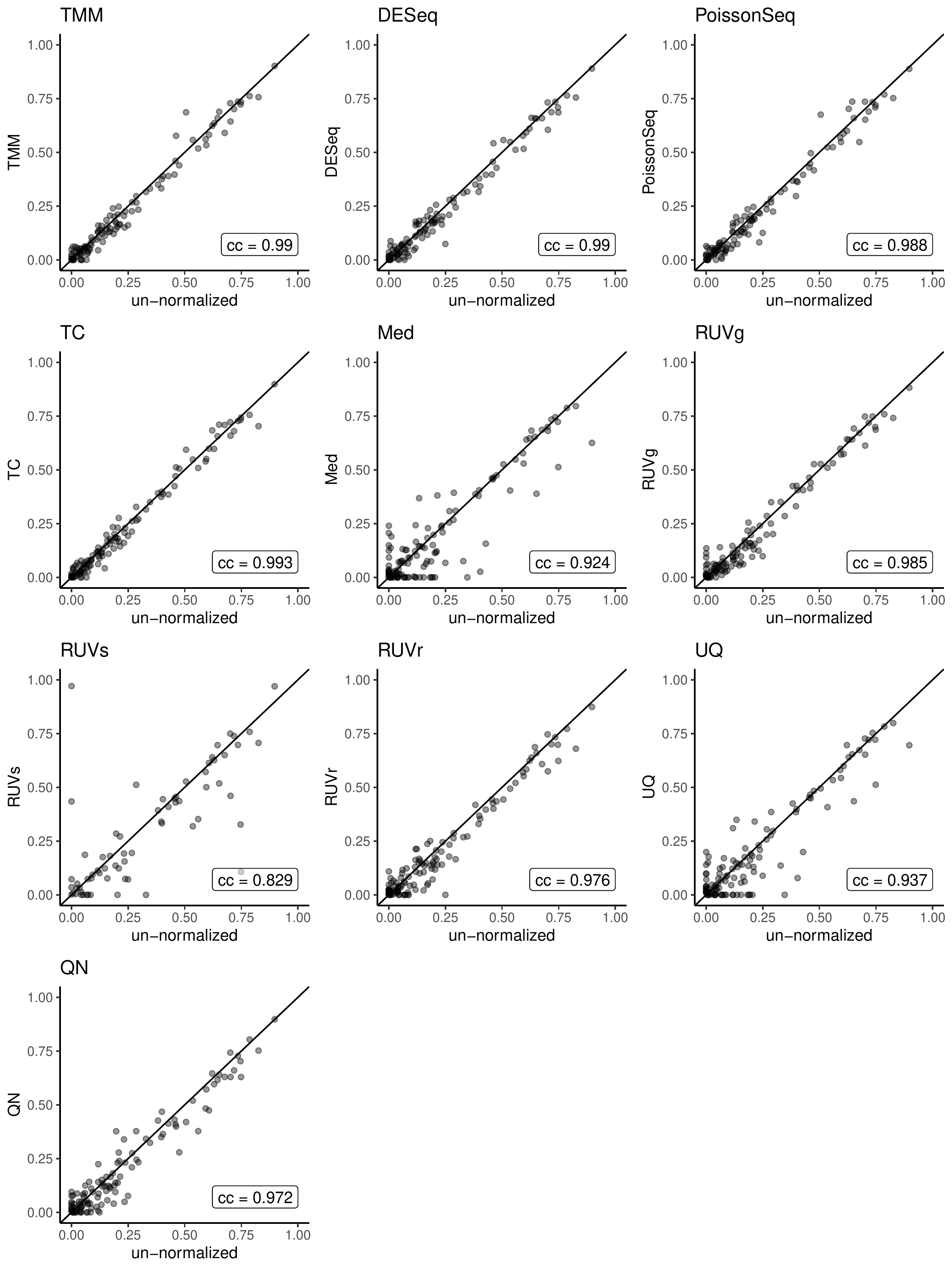} 
	\caption{
		Scatter plots of partial correlations in positive controls before versus after normalization for each normalization method for the TCGA-BRCA/UCS data.
		The corresponding concordance correlation coefficient between the shown partial correlations, which is measured by~\cc, is indicated in the lower right corner of each plot.
	}
	\label{fig:BRCA-UCS-pcorr-scatter-all}
\end{figure}

For positive controls, we again observe a high abundance of within-cluster correlations as expected for clustered miRNAs (Fig.~\ref{fig:BRCA-UCS-pcorr-all-norm-methods}).
Normalization can alleviate off-cluster correlations in terms of both the number and strength.
Finally, depending on the method, normalization can either retain or reduce within-cluster correlations, which is signified by their before-versus-after-normalization concordance (Fig.~\ref{fig:BRCA-UCS-pcorr-scatter-all}).

\clearpage

\bibliographystyle{apalike}
\bibliography{library}